\documentclass[11pt,a4paper,oneside]{article}

\usepackage[top=3cm, bottom=3cm, left=2cm, right=2cm]{geometry}
\usepackage[english]{babel}
\usepackage[utf8]{inputenc}
\usepackage[T1]{fontenc}

\usepackage{amsthm,amsmath,amssymb,mathrsfs,dsfont,mathtools, amsfonts}
\usepackage[hidelinks]{hyperref}
\usepackage[noabbrev, capitalise]{cleveref}
\usepackage{framed}
\usepackage{bbm}
\usepackage{bm}
\usepackage{graphicx}
\usepackage{xcolor}
\usepackage[ruled]{algorithm2e}
\usepackage{algorithmic, setspace}
\usepackage[shortlabels]{enumitem}
\usepackage{verbatim}
\usepackage{float}

\usepackage[authoryear, round]{natbib}
\usepackage{authblk}
\date{}
\title{Infinite joint species distribution models}

\author[1]{Federica Stolf}
\author[1]{David B. Dunson}
\affil[1]{Department of Statistical Science, Duke University, Durham, NC, USA}

\providecommand{\keywords}[1]{
  \small	
  \textbf{\textit{Keywords:}} #1
  \normalsize
}

\usepackage[sf]{titlesec}
\usepackage{sectsty}
\allsectionsfont{\centering \normalfont\scshape} 

\newtheorem{theorem}{Theorem}

\newtheorem{lemma}{Lemma}
\newtheorem{proposition}{Proposition}
\theoremstyle{definition}

\newtheorem{definition}{Definition}

\newcommand{\commons}{c^{(\epsilon)}}

\newcommand{\V}{\mathrm{var}}
\newcommand{\R}{\mathds{R}}
\newcommand{\pr}{\mathrm{pr}}
\newcommand{\cov}{\mathrm{cov}}

\begin{document}

\maketitle

\begin{abstract}
Joint species distribution models are popular in ecology for modeling covariate effects on species occurrence, while characterizing cross-species dependence. Data consist of multivariate binary indicators of the occurrences of different species in each sample, along with sample-specific covariates. A key problem is that current models implicitly assume that the list of species under consideration is predefined and finite, while for highly diverse groups of organisms, it is impossible to anticipate which species will be observed in a study, and discovery of unknown species is common. This article proposes a new modeling paradigm for statistical ecology, which generalizes traditional multivariate probit models to accommodate large numbers of rare species and new species discovery. We discuss theoretical properties of the proposed modeling paradigm and implement efficient algorithms for posterior computation. Simulation studies and applications to fungal biodiversity data provide compelling support for the new modeling class.
\end{abstract}


\section{Introduction}
Joint species distribution models are used routinely in ecology. The focus is on modeling covariate effects and cross-species dependence in occurrence. Current models implicitly assume that a list of the $p$ species under study can be prespecified. Letting $j \in \{1,\ldots,p\}$ index the species in this list, species co-occurrence data consist of an $n \times p$ matrix $Y$ of binary indicators with $y_{ij}=1$ if species $j$ was found in sample $i$ and $y_{ij}=0$ otherwise. Joint species distribution models define a multivariate binary regression for $y_i = (y_{i1},\ldots,y_{ip})^{\top}$ given covariates $x_i = (x_{i1},\ldots,x_{iq})^{\top}$. For example, the popular \texttt{HMSC} package relies on a Bayesian probit latent factor model \citep{OvaskainenEtal2016,TikhonovHmsc20200}. 

For common species the above framework is highly successful, providing key insights into the role of biotic and abiotic factors on community assembly. However, with the emergence of automated biodiversity monitoring methods for highly diverse groups of organisms, such as arthropods and fungi, critical problems emerge. First, while biomass is dominated by a relatively small number of common species, the vast majority of species on Earth are extremely rare. Hence, it is impossible to pre-specify a list of species under study in advance of collecting data. In a given study, we tend to observe a tiny fraction of species known to occupy the study region. In addition, it is common to discover species unknown to science. Given these problems, current practice chooses the list of species to be considered in the joint species analysis in a post hoc manner, typically discarding the rare species that make up most of the biodiversity.

\begin{figure}
\begin{center}
\includegraphics[width=0.5\textwidth]{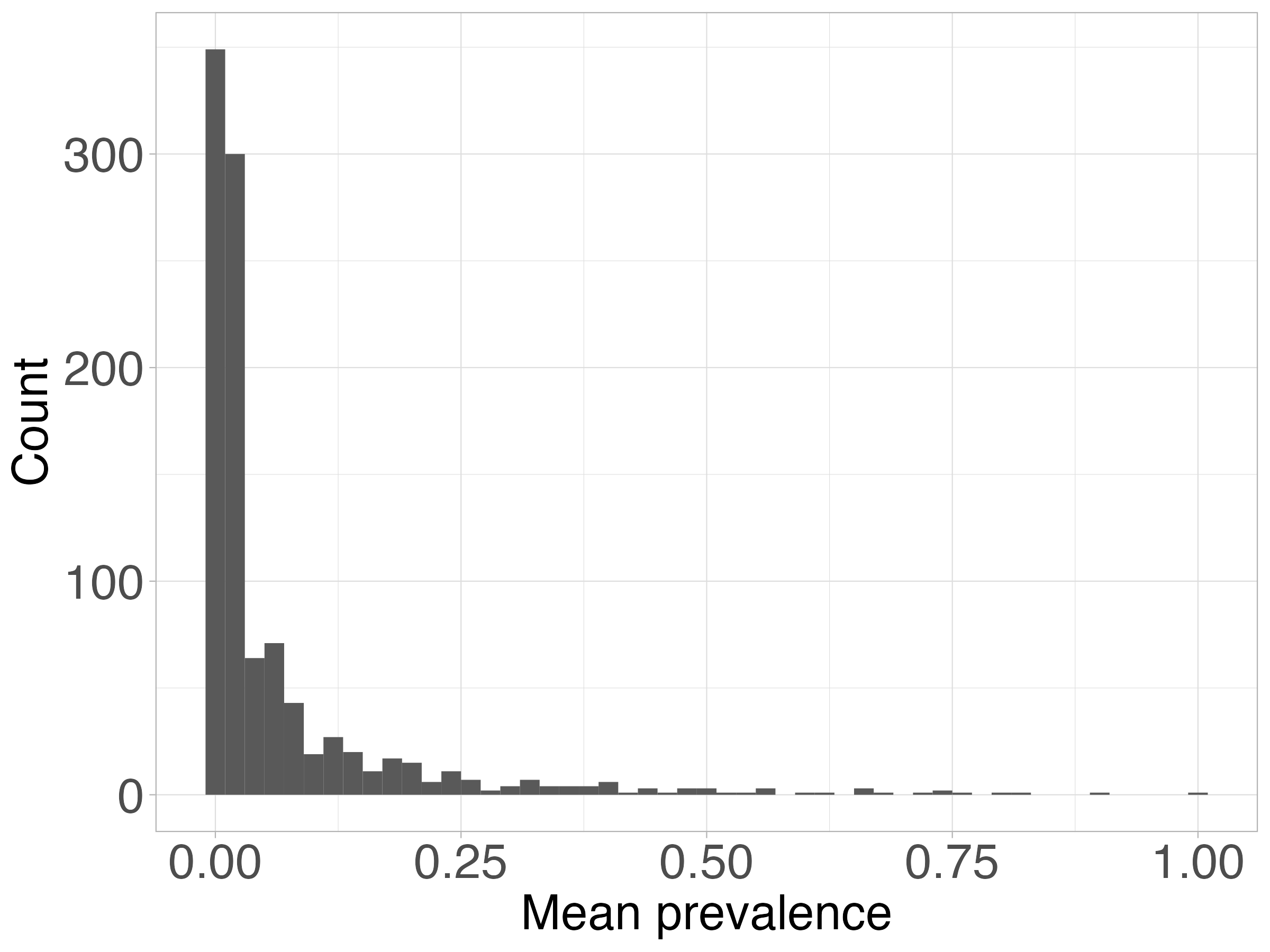}
\caption{Empirical species occurrence probabilities for fungi data in a Finnish study.}
    \label{fig:fungi_prev}
    \end{center}
\end{figure}

To make this motivation concrete, in Figure \ref{fig:fungi_prev} we plot the empirical probabilities of the occurrence of fungi species from \cite{AbregoFungi}. Although we use the word species for simplicity, ecologists instead refer to operational taxonomic units (\texttt{OTU}). The \texttt{OTU}s are obtained by applying DNA meta-barcoding to the samples and then implementing a taxonomic classifier \citep{somervuo2016unbiased}.
The proportion of DNA sequences classified to each OTU does not provide a reliable measure of relative OTU abundance within each sample due to vast differences in size (spore size for fungi, body size for arthropods). Hence, it is standard practice in ecology to use meta-barcoding only to produce indicators of the presence or absence of each OTU in each sample. Inferring relative abundances of different organisms from relative read abundances remains one of the major open challenges for DNA-based approaches \citep{compson2020metabarcoding}.
Similar data on fungi co-occurrence were the focus of a recent {\em Nature} paper \citep{abrego2024airborne}; the authors analyzed the data using the \texttt{HMSC} package but focusing on the few 100 most common species and discarding $\sim 20,000$ rare species.
However, rare species have a critical impact on biodiversity, so ignoring them limits the conclusions that can be drawn from the analysis.

To motivate directions forward in addressing these problems, consider the multivariate probit model  \citep{ChibMultivariateProbit}: 
\begin{eqnarray}
y_{ij} &= & \mathds{1}(z_{ij}>0),\quad  z_i = (z_{i1},\ldots,z_{ip})^{\top} \nonumber \\
z_i & = & B^{\top} x_i + \varepsilon_i,\quad 
\varepsilon_i \sim N_p(0,\Sigma),\label{eq:mvp}
\end{eqnarray}
with $B = (\beta_1, \dots, \beta_p)$ a $q \times p$ matrix of regression coefficients, $\Sigma$ a correlation matrix and $\mathds{1}(\cdot)$ the indicator function. Under this model, the marginal probability of occurrence of species $j$ conditionally on covariates $x_i=x$ is $\pi_j = \Phi(x^{\top}\beta_j)$, with $\Phi(\cdot)$ the standard normal cumulative distribution function. Dependence in co-occurrence between species $j$ and $j'$ is captured by $\mbox{cor}(z_{ij},z_{ij'})=\sigma_{jj'}$, the $j,j'$ element of matrix $\Sigma$. 

In implementing Bayesian analyses of fungi biodiversity data using model \eqref{eq:mvp}, there are fundamental problems. The first is that, as described above, we cannot tie down the identities of the different binary outcomes (species) $j=1,\ldots,p$ in advance. In fact, we expect to regularly discover new species as we collect additional samples, so that the total number of species in $n$ samples may be substantially less than that in $n+m$ samples. Hence, we need a mechanism to allow for a growing number of binary outcomes. 

A related problem is considered in the literature on infinite latent feature models and the Indian Buffet Process (\texttt{IBP}) \citep{JMLR:v12:griffiths11a}. 
The \texttt{IBP} provides a generative model for binary matrices having infinitely many columns, derived as the limiting case of a beta-Bernoulli process that samples the elements independently from Bernoulli distributions with column-specific probabilities drawn independently from Beta distributions. The \texttt{IBP} allows new binary features (species) to be discovered as new samples are collected, but does not characterize within-sample dependence in species occurrences. Moreover, the \texttt{IBP} does not accommodate  covariates, limiting not just predictive performance but also interpretability to ecologists, who are interested in how species communities and biodiversity differ with various factors including climate, habitat and human footprint.
Potentially, we could use model \eqref{eq:mvp} setting $p$ to a large number with a buffer incorporated to allow new species to be added as the sample size increases. However, simply adding a buffer without careful consideration is ad hoc and may have unanticipated negative consequences, as we will show later in the manuscript. 

Motivated by the above considerations, the focus of this article is on developing a general new infinite joint species distribution modeling framework, borrowing key ideas from the literature on infinite latent feature models to appropriately extend multivariate probit-style hierarchical regression models. The proposed framework avoids discarding rare species in inferring covariate effects on species communities and in investigating cross-species dependence. In addition,
a key advantage over traditional fixed species list models is that our framework enables prediction of the number of additional species that would be discovered if $m$ additional samples were collected after an initial $n$ samples. Such predictions are of substantial interest for making decisions about whether to collect additional data in a biodiversity monitoring study.  

Our framework adds to the ongoing literature on generalizations of the classical \texttt{IBP} model to allow additional flexibility; refer, for example, to \cite{Broderick2013FeatureAP,  gershman2014distance, di2020non, camerlenghi2022scaled, williamson2010dependent,  warr2022attraction}. In this regard, we are the first to incorporate dependence within samples in feature occurrence.

\section{Proposed Modeling Framework}

\subsection{Model specification}

We introduce a novel modelling framework, which we call TRACE (tracking rare and abundant co-occurrences in ecology).
Let $Y$ be a binary matrix of species occurrences with $n$ rows and an unbounded number of columns, where $y_{ij}=1$ indicates that species $j$ was found in the sample $i$. The matrix $Y$ is sampled as a sequential process to accommodate a growing number of species. In this way, we avoid fixing species identities in advance, but let the model adapt to the growing complexity of the data as new species are discovered. As commonly done in infinite latent feature models, we derive a distribution on infinite binary matrices by starting with a truncation level $p$ for the number of species and then taking the limit as $p \rightarrow \infty$. 

Our proposed infinite joint species distribution model is defined through a latent Gaussian construction. Let ${z_i=(z_{i1}, \dots, z_{ip})^{\top} \in \R^{p}}$ denote a $p$-variate latent continuous variable for $i=1,\dots,n$ and let $y_{ij}$ be 1 or 0 according to the sign of $z_{ij}$.
Let $x_i=(x_{i1}, \dots, x_{iq})^{\top}$ be the covariate for each sample $i$ and $\beta_j = (\beta_{j1}, \dots, \beta_{jq})^{\top}$ the regression coefficients specific to species $j$, with $q$ the number of covariates. 
The \texttt{TRACE} modelling framework is defined as
\begin{align} \label{eq:TRACE}
    &y_{ij} = \mathds{1}(z_{ij}>0), \quad z_{ij} = \alpha_j + x_i^{\top} \beta_j + \varepsilon_{ij},  
    \\ \nonumber
       & \alpha_j \sim N(\mu_p, \tau_p^2),\quad \varepsilon_i \sim N_p(0,\Sigma), 
\end{align} 
where $\alpha=(\alpha_1, \dots, \alpha_p)^{\top}$ is a $p$-dimensional vector of random intercepts, with $\alpha_j$ controlling the baseline commonness of species $j$, and $\Sigma = \{ \sigma_{jj'} \}$ is a  $p \times p$ positive definite correlation matrix, with $\sigma_{jj'}$ controlling dependence in occurrence of species $j$ and $j'$. We will later describe hierarchical models for $B=(\beta_1,\ldots,\beta_p)$ and $\Sigma$ to allow the borrowing of information. 

The hierarchical structure on the intercepts $\alpha_j$ plays a critical role in the model specification. In order to accurately characterize the data, we need to allow there to be a small proportion of common species and a large number of very rare species. In addition, in the limit as $p \to \infty$, the number of species observed in a sample $\sum_{j=1}^p y_{ij}$ should be almost surely finite. The hyperparameters $\mu_p$ and $\tau_p$ in the prior for $\alpha_j$ play a crucial role in determining the induced likelihood of $z_i$ and thus also the likelihood of $Y$. These hyperparameters control the distribution of the number of species within each sample, as well as the rate of growth in the total number of species, corresponding to the number of non-zero columns of $Y$ as $n$ increases.

We will show in the next section that the default values
\begin{equation} \label{eq:mu_tau}
    \mu_p = (1+ \tau_p^2)^{1/2}\,\Phi^{-1}\bigg(\frac{\gamma}{\gamma + p}\bigg), \quad \tau_p = (2 \log p)^{1/2},\quad 
    \gamma \in \mathbb{R}^{+}
\end{equation}
lead to several desirable properties as $p$ grows. In the baseline case in which there are no covariates and $\Sigma = I$, \texttt{TRACE} is shown to have a similar behavior to the popular Indian buffet process. Hence, \texttt{TRACE} can be viewed as providing a useful framework for extending the \texttt{IBP} to incorporate within-sample and covariate dependence in feature occurrences.

The \texttt{TRACE} framework corresponds to a probit-Bernoulli process: it generates species from a Bernoulli process with probabilities drawn from an underlying probit process. Hence, marginally for each of the species we have $\pr(y_{ij}=1 \mid \beta_j, \alpha_j, x_i) =  \Phi(\alpha_j + x_i^{\top}\beta_j)$.
For rare species, there is insufficient information in the data to accurately estimate species-specific regression coefficients. To address this problem, it is natural to choose a hierarchical structure to borrow information across the different species. Hence, we let
\begin{equation} \label{eq:b_hier}
    \beta_j \mid \nu, \Psi \sim N_q(\nu, \Psi), \quad (\nu, \Psi) \sim \mathrm{NIW}(\nu_0,\iota,d,\Xi),
\end{equation}
where $\mathrm{NIW}(\nu_0,\iota,d,\Xi)$ is a Normal-Inverse Wishart prior distribution. 

Similarly to the usual multivariate probit model \eqref{eq:mvp}, the likelihood of $Y$ under the \texttt{TRACE} model cannot be expressed in a closed form, as it requires integrating over a multivariate normal distribution. 
However, there is a rich literature on efficient posterior computation algorithms for \eqref{eq:mvp}, which we will leverage. We demonstrate several compelling theoretical properties that provide strong support for \texttt{TRACE}. For example, we will show that the limiting number of non-zero entries in $Y$ as $p \rightarrow \infty$ is almost surely finite. This ensures that the number of species sampled will be finite even in the limiting case.

\subsection{Properties and connection to Indian buffet processes} \label{sec:theory}

In this section, we present some theoretical properties motivating the modeling framework described in (\ref{eq:TRACE}) and (\ref{eq:mu_tau}). Proofs are given in the supplementary materials. We also highlight relationships with the Indian buffet process. To emphasize this connection, we start by considering the \texttt{TRACE} model without covariates: 
\begin{equation} \label{eq:interceptTRACE}
    y_{ij} = \mathds{1}(z_{ij}>0), \quad z_{ij} = \alpha_j  + \varepsilon_{ij}, \quad \alpha_j \sim N(\mu_p, \tau_p^2), \quad \varepsilon_i \sim N_p(0,\Sigma)
\end{equation} 
with the values for $\mu_p$ and $\tau_p$ in equation \eqref{eq:mu_tau}. We recall that the Indian buffet process is induced by a Beta-Bernoulli process: $y_{ij} \mid \pi_j \sim \mathrm{Bernoulli}(\pi_j),$  $\pi_j \sim \mathrm{Beta}(\gamma/p,1)$, where the $\pi_j$s and $y_{ij}$s are mutually independent
\citep{thibaux2007hierarchical}.

First, we prove that in the limiting case as $p\to \infty$, the expected number of non-zero entries is finite. More specifically, Theorem \ref{thm:finite_ni} shows that the no covariates \texttt{TRACE} model has the same number of expected features as the \texttt{IBP}, yielding an intuitive interpretation for the parameter $\gamma$. It also shows that in the simple case with $\Sigma = I$, where there is no dependence between species as in the \texttt{IBP}, the number of species per sample has the same distribution as the \texttt{IBP}.

\begin{theorem} \label{thm:finite_ni}
    Consider the no covariates \texttt{TRACE} model and let $n_{i+}= \sum_{j=1}^p y_{ij}$.
    Then $\lim_{p \to \infty} E(n_{i+}) = \gamma,$
    for each $i=1,\dots,n$. If we further assume $\Sigma = I$, then $n_{i+} \rightarrow Poi(\gamma)$ in distribution as $p \to \infty$.
\end{theorem}

Theorem \ref{thm:finite_ni} ensures that the number of species sampled is finite, so $\pr(n_{i+} < \infty)=1$.
In the following proposition, we characterize the limiting variance of $n_{i+}$ in the no covariates $\texttt{TRACE}$ model.

\begin{proposition} \label{prop:variance_ni}
    Consider the no covariates TRACE model and assume that $\sigma_{jj'} \in \{\rho_1,\ldots,\rho_k\}$ with $k \in \mathds{N}$. Then $ \gamma + \gamma^2/2 \, \{\underset{k}{\mathrm{min}}\, |\exp(\rho_k) -1|\} \le \lim_{p \to \infty} \V(n_{i+}) \le \gamma + \gamma^2/2 \, \{\underset{k}{\mathrm{max}} \, |\exp(\rho_k)-1|\}$, for each $i=1,\dots,n$.
\end{proposition}

Proposition \ref{prop:variance_ni} provides bounds for the limiting variance of $n_{i+}$. For tractability in deriving and interpreting the bounds, we make the simplifying assumption that the between-species correlations belong to a finite set of possible values. If some pairs of species are uncorrelated, the lower bound is $\gamma$, while if some pairs of species are perfectly correlated, the upper bound is
$\gamma + \gamma^2\{\exp(1)-1\}/2$. These extreme case bounds can be shown to hold even when the finite set assumption is violated. In the $\Sigma=I$ case the limiting variance is $\gamma$. 
The above results are extended to the general \texttt{TRACE} model, which allows inclusion of covariates, in Theorem \ref{thm:cov_ni}. 

\begin{theorem} \label{thm:cov_ni}
Consider the \texttt{TRACE} model and let $g_i = \exp\{ x_i^{\top}\nu + (1/2) x_i^{\top}\Psi x_i\}$. Then for $i=1,\dots n$,  
    (i) $\lim_{p \to \infty} E(n_{i+})=\gamma g_i$;
    (ii)  if we further assume  $\sigma_{jj'} \in \{\rho_1,\ldots,\rho_k\}$ with $k \in \mathds{N}$, we have $\gamma g_i(1+  \gamma g_i/2) \, \{\underset{k}{\mathrm{min}}\, |\exp(\rho_k) -1|\} \le \lim_{p \to \infty} \V(n_{i+}) \le \gamma g_i(1+\gamma g_i/2)\, \{\underset{k}{\mathrm{max}} \, |\exp(\rho_k)-1|\}$.
\end{theorem}

The above theorem provides valuable insights into the impact of covariates on biodiversity measured in terms of sample-specific species richness, that is, the number of species per sample.
The expected sample-specific species richness is simply $\gamma g_i$, where $\log (g_i)$ follows a quadratic regression in covariates $x_i$ with linear coefficients corresponding to the mean $\nu=E(\beta_j)$ and quadratic coefficients to the covariance $\Psi=\mbox{cov}(\beta_j)$ of the population distribution of species-specific coefficients in \eqref{eq:b_hier}. Obtaining
such a simple induced model characterizing the impact of covariates on biodiversity is a major advantage of \texttt{TRACE}. For example, the impact of the covariates of climate or environmental disturbances on the richness of sample-specific species can be directly assessed using Theorem
\ref{thm:cov_ni}.

A critical property of biodiversity data is that we expect a small number of common species having $\pi_j$ values not close to zero with the remaining having $\pi_j \approx 0$. This leads to considerable sparsity in the $Y$ data matrix with many columns having few 1s. The following proposition supports the sparsity of the \texttt{TRACE} model.

\begin{proposition} \label{prop:sparsity}
The data matrix $Y$ sampled under the \texttt{TRACE} model is sparse in the sense that 
$\lim_{p \to \infty} \pr(y_{ij}=1)  =0$ and $\lim_{p \to \infty} p\, \pr(y_{ij}=1)/(\gamma g_i)  =1$ for each $i=1,\dots,n$.
\end{proposition}
Proposition \ref{prop:sparsity} shows that as $p$ increases the marginal probabilities $\pr(y_{ij}=1) = o(1)$, where the notation $f(p)=o(1)$ means that $\lim_{p\to \infty} f(p)/c=0$ for any constant $c$. The same sparsity property also holds for the \texttt{IBP}, as for the \texttt{IBP} one has $\pr(y_{ij}=1) = o(1)$ as $p\to \infty$. 

In addition to the sparsity of $Y$, a key aspect of the data is that some species are common. Such species have $\pi_j$ values in the right tail of the population distribution of $\pi_j$ across $j$. In order for the model to be realistic in biodiversity contexts, we need the sparsity property of Proposition \ref{prop:sparsity} to not rule out the presence of common species. Thus, the focus is on studying the right tail of the distribution of $\pi_j$s across species. In definition \ref{def:common}, we provide a formal notion of common species.

\begin{definition} \label{def:common}
Let $Y$ be an $n \times p$ binary matrix of species occurrence indicators and let
$\pi_j=\pr(y_{ij}=1)$ for $i=1, \dots,n$. Species $j$ is common if $\pi_j > \epsilon$ for a fixed threshold $\epsilon>0$.
\end{definition}

Let $c_j^{(\epsilon)} = \mathds{1}(\pi_j > \epsilon)$ indicate whether the $j$th species is common, for $j=1,\dots,p$, 
and $\commons = \sum_{j=1}^p c_j^{(\epsilon)}$ denote the number of common species for the threshold $\epsilon$. The latter quantity summarizes the behaviour of the marginal probability distribution in $j$ and characterizes the right tail of this distribution for high values of $\epsilon$.

For the \texttt{IBP} model, the expected number of common species $ E(\commons)$ as $p$ grows is
\begin{equation*}
    \lim_{p\to \infty} \sum_{j=1}^p \pr(\pi_j > \epsilon) = \lim_{p\to \infty} p\{1- B_{\epsilon}(\gamma/p,1)\}= \lim_{p\to \infty} p(1-\epsilon^{\gamma/p}) = -\gamma \log \epsilon,
\end{equation*}
where $B_x(a,b)$ is the regularized incomplete beta function. This is a novel result to our knowledge. The following proposition states a related result for the no covariates \texttt{TRACE} model.

\begin{proposition} \label{prop:commonf}
For no covariates \texttt{TRACE} model the expected number of common species is $\lim_{p \to \infty} E(\commons) \\ = \gamma  \exp\{-\Phi^{-1}(\epsilon) -0.5\}$.
\end{proposition} 

\begin{figure}
\begin{center}
\includegraphics[width=0.6\textwidth]{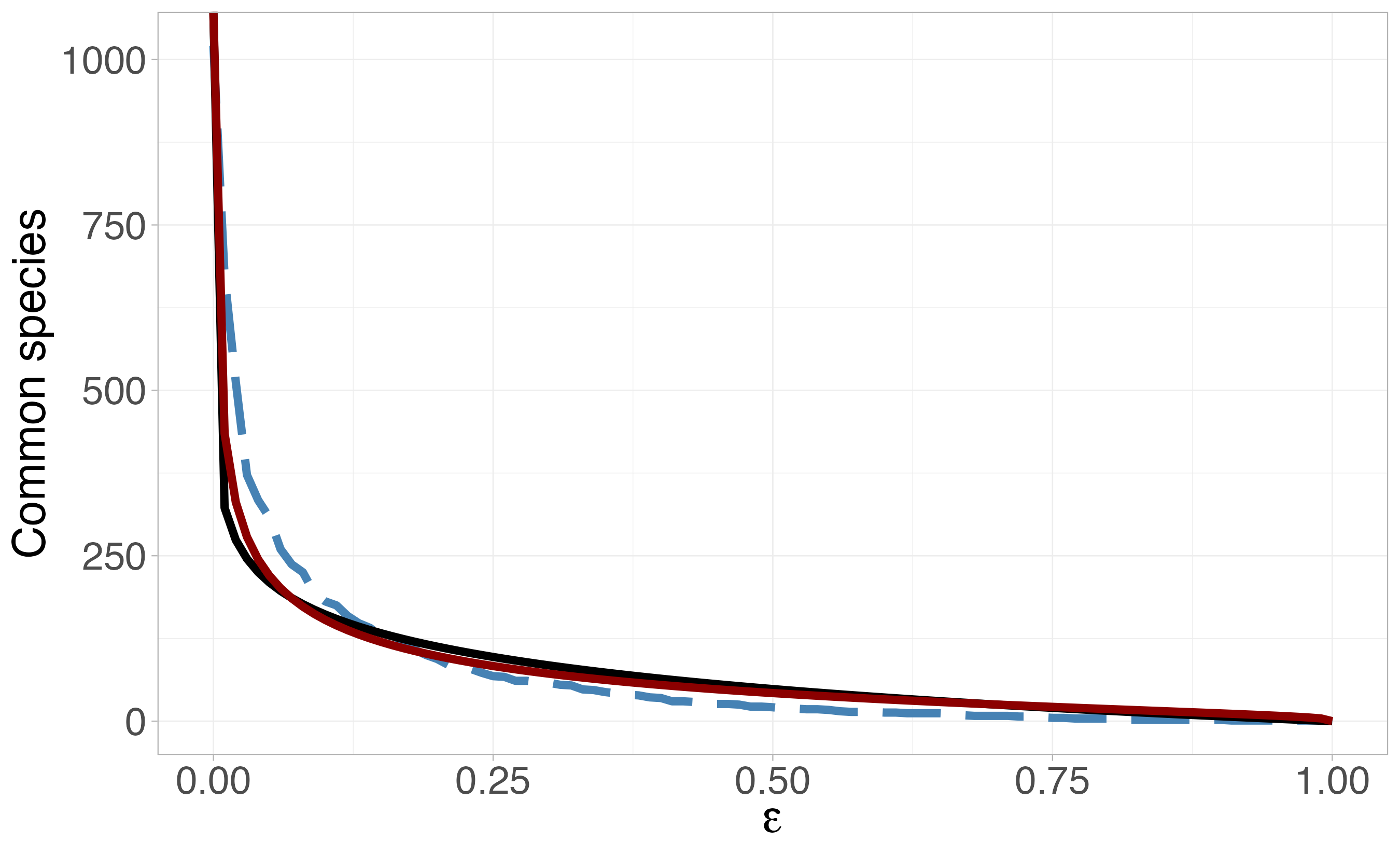}
\caption{Expected total number of common species for the \texttt{TRACE} (red line) and \texttt{IBP} (black line) models for different values of $\epsilon$, with $\gamma = 70$. The dashed blue line shows the empirical number of common species for a grid of values of $\epsilon$ for the fungi co-occurrence data.}
    \label{fig:comparison_common}
    \end{center}
\end{figure}

Hence, the expected number of common species under \texttt{TRACE} differs somewhat from that of \texttt{IBP}. However, the functional relationship with the commonness threshold $\epsilon$ is similar, as illustrated in Figure \ref{fig:comparison_common}.
 This figure also compares the expected number of common species under the \texttt{TRACE} and \texttt{IBP} models with the empirical values observed in our motivating data. For this plot, we used an optimized value of $\gamma$, set to 70. The theoretical curves for the \texttt{TRACE} and \texttt{IBP} models align closely, indicating that both models exhibit comparable behavior. Moreover, they effectively capture the observed total number of common species in the fungi co-occurrence data. The following proposition adapts the result to the general \texttt{TRACE} framework. 

\begin{proposition} \label{prop:cov_common}
Let $c_{ij}^{(\epsilon)} = \mathds{1}\{\Phi(\alpha_j + x_i^{\top}\beta_j) > \epsilon\}$ and $c_i^{(\epsilon)} = \sum_{j=1}^p c_{ij}^{(\epsilon)}$ denote the number of common species with threshold $\epsilon$ for the sample $i$ ($i=1,\ldots,n)$.
For \texttt{TRACE}, the expected number of common species for sample $i$ is 
     \begin{equation*}
        \lim_{p \to \infty} E(c_i^{(\epsilon)}) = \gamma  \exp\Big\{-\Phi^{-1}(\epsilon) +
        x_i^{\top}\nu +
        (1/2)x_i^{\top}\Psi x_i -0.5 \Big\} = \gamma g_i 
        \exp\{ -\Phi^{-1}(\epsilon) -0.5\}.
    \end{equation*}
\end{proposition} 
As in Theorem \ref{thm:cov_ni} for the expected species richness of the sample, we obtain a quadratic regression model for the log of the expected number of common species. Interestingly, the expected number of common species in a sample is the expected species richness multiplied by $\exp\{ -\Phi^{-1}(\epsilon)-0.5 \}$.

In biodiversity studies, the total number of species observed in a study, or overall species richness, is of key interest. This quantity corresponds to the number of non-zero columns of $Y$ observed in $n$ samples, which we denote as $p_n^*$. 
\cite{JMLR:v12:griffiths11a} showed that the expected value of $p^*_n$ for the \texttt{IBP} is $\gamma H_n$, where $H_n$ is the $n$th harmonic number.
Theorem \ref{prop:pstar} characterizes the growth rate for the no covariates \texttt{TRACE} model, while providing further insights into the expected value and distribution of $p_n^*$ in the simple case with $\Sigma = I$. 

\begin{theorem} \label{prop:pstar} 
Consider the no covariates \texttt{TRACE} model and let $w_j = \mathds{1}(n_{+j}>0)$ with $n_{+j}=\sum_{i=1}^n y_{ij}$ and $p^*_n = \sum_{j=1}^p w_j$. Then $\lim_{p \to \infty} E(p^*_n)=\gamma \exp(-1/2) \sqrt{2\pi}C_n$, with $C_n= \int_0^1 \{1-(1-t^n)\}\exp\{-\Phi^{-1}(t)\\\}  /\exp[-0.5\{\Phi^{-1}(t)\}^2] dt$. If $\Sigma = I$, then $p^*_n \rightarrow \mathrm{Poi}\{\gamma\exp(-1/2) \sqrt{2\pi}C_n\}$ in distribution as $p \to \infty$.
\end{theorem} 
Figure S1 in the supplementary material shows the empirical behavior of $C_n$ for different values of $n$, comparing it to $H_n$ and other functions of $n$.

\subsection{Correlation structure and prior elicitation} \label{sec:corr}

The \texttt{TRACE} model defined in \eqref{eq:TRACE} and \eqref{eq:mu_tau} provides a general framework that can be used in different settings with appropriate choices of the correlation matrix. Here, we propose alternative approaches to model $\Sigma$ for reducing dimensionality and overcoming the inferential challenges of rare species. Defining a structure that induces shrinkage in the correlation matrix is essential considering the limited data available for accurately estimating its elements.

We consider three different models for the correlation matrix $\Sigma$. The first assumes a common correlation coefficient for all pairs of species, $\rho_{jj'} = \rho$. We induce a prior distribution for $\rho$ by a Gaussian prior for the Fisher transformation of $\rho$: $\zeta =0.5 \log\{(1+\rho)/(1-\rho)\}$ with $\zeta \sim N(0, w_0^2)$. 
In addition to being a simple dependence structure, this formulation effectively reduces the dimensionality of $\Sigma$ and may be realistic in certain contexts. For example, a common positive correlation can be induced by including a random intercept $\xi_i\sim N(0,\lambda)$ in \eqref{eq:TRACE} that measures how beneficial the conditions are in general to all the different species under study.

As a more flexible alternative, we consider the popular Lewandowski-Kurowicka-Joe (\texttt{LKJ}) prior \citep{lewandowski2009generating} for $\Sigma$, which is proportional to $|\Sigma|^{\nu-1}, \nu > 0.$ For $\nu=1,$ the prior is uniform over the set of correlation matrices. However, for a large number of sparsely observed species, borrowing information across species through a hierarchical model is important. Hence, we also consider a hierarchical shrinkage prior that lets
\begin{equation}   \label{eq:corrH}
   \zeta_{jj'} = 0.5 \log \{(1+\sigma_{jj'})/(1-\sigma_{jj'})\}, \quad \zeta_{jj'} \mid \omega^2  \sim N(0, \omega^2), \quad \omega^2 \sim IG(a_{\omega},b_{\omega}),  
\end{equation}
where $IG(a,b)$ is an inverse-gamma distribution. We also tried a more flexible variation, which modelled the 
$\zeta_{jj'}$ distribution as Gaussian with an unknown mean and variance, but this did not improve performance so we focus on the simpler zero-mean form.
In the supplementary materials, we show the marginal prior density for $\sigma_{jj'}$ induced from \eqref{eq:corrH} and provide a figure that plots this density. In practice, we observe dramatically better performance for the hierarchical shrinkage prior \eqref{eq:corrH} than for the \texttt{LKJ} prior.

To complete a Bayesian specification of the \texttt{TRACE} model, we choose a gamma prior distribution for the parameter $\gamma$, $\gamma \sim Ga(a_{\gamma},b_{\gamma})$.

\subsection{Posterior computation} \label{sec:post_comp}

Posterior inference is carried out by truncating the number of species to a suitable value $p$, which is substantially greater than the number of species observed $p_n^*$. Truncation is standard practice in nonparametric Bayesian models involving infinitely many components, ranging from mixture models \citep{ishwaran2001gibbs} to latent feature models \citep{williamson2010dependent, doshi2009variational}. In the supplementary materials, we provide simulation results showing that posterior inferences for the \texttt{TRACE} model are robust with respect to the choice of truncation level.

Fitting \texttt{TRACE} presents the key computational challenge of evaluating multivariate Gaussian orthant probabilities. A common strategy in fitting multivariate probit models is the use of data augmentation with latent variables simulated from truncated multivariate Gaussian distributions \citep{ChibMultivariateProbit}.
However, this approach is impractical for \texttt{TRACE}.
Sampling from high-dimensional truncated Gaussian distributions is computationally prohibitive and remains an active research area. Additionally, 
for imbalanced binary data, \cite{JohndrowImbalanced} showed that Markov chain Monte Carlo algorithms that rely on data augmentation suffer from poor mixing. The large number of rare species results in a highly imbalanced setting, making the mixing issue particularly problematic. 

To facilitate scaling to large numbers of species, one possibility is to adapt a recently proposed algorithm for rapid posterior computation in multivariate probit models \citep{chakraborty2022bayesian}.
This algorithm focuses on approximating marginal posterior distributions for the parameters, obtaining accurate point estimation and uncertainty quantification. It is built on a Laplace approximation strategy \citep{tierney1986accurate}.
The detailed steps of the algorithm for posterior inference for \texttt{TRACE} are given in Section S.3 of the supplementary materials. By focusing on marginal posterior approximations, marginal priors for $\sigma_{jj'}$ can be used that do not explicitly enforce the positive semidefinite constraint on $\Sigma$ \citep{chakraborty2022bayesian}.

\section{Performance assessments in simulations} \label{sec:simulation}

We evaluated the performance of \texttt{TRACE} through simulation studies. One metric is accuracy of estimation relative to the true marginal occurrence probabilities $\pi_{ij}^* = \Phi(\alpha_j^* + x_i^{\top}\beta_j^*)$ and the true correlation matrix $\Sigma^*$. In ecological applications, there is substantial interest in predicting the number of new species discovered in an additional $m$ samples, conditionally on all the data.  We use $\Delta_{m|n} = p^*_{n+m} - p^*_n$ as the notation for this quantity. We will focus on predicting the number of new species in the last 20 samples, the key quantity being $\Delta_{100|80}^* =  p^*_{100} - p_{80}^*$.

We compare with the \texttt{IBP} and a multivariate probit model with a data matrix augmented with zeros for all columns $j$ such that $p_n^* < j\le p$.
Inference in the \texttt{IBP} model proceeds through a Gibbs sampler, considering a gamma prior for $\gamma$ and following the truncated beta-binomial representation of \cite{JMLR:v12:griffiths11a}. For the \texttt{IBP} and \texttt{TRACE} models, we set the parameters for the gamma prior on $\gamma$ such that the distribution is centered on the empirical mean of the number of species per sample. For \texttt{TRACE} we use the hierarchical prior on $\Sigma$ in \eqref{eq:corrH} or the \texttt{LKJ} prior; results for the \texttt{LKJ} prior were significantly worse, as shown in Figure \ref{fig:lkj} of the supplementary material.
The multivariate probit model is estimated using the \texttt{bigMVP} R package, which implements the fast posterior algorithm discussed in \cite{chakraborty2022bayesian}, assuming a hierarchical normal prior on the regression coefficients and prior \eqref{eq:corrH} for 
the correlation matrix. 

To simulate the data, we choose a sample size of $n=100$ and $q=5$ covariates and vary the number of observed species $p^*_n$, which is upper bounded by $p=500$. The parameter $\gamma$ is varied, ranging from 1 to 20, obtaining 20 different simulated datasets. As $\gamma$ increases, more species are sampled, as described in Section \ref{sec:theory}. For each $\gamma$, we consider three data-generating scenarios: 1) (factor) data are simulated under the \texttt{TRACE} model in \eqref{eq:TRACE} and \eqref{eq:mu_tau} with $\Sigma^* = \Lambda \Lambda^{\top} + I$ where $\Lambda$ is a $p\times H$ matrix with $H=50$ and $\lambda_{jh} \sim N(0,1)$,  
2) (tobit) data are generated under a misspecified \texttt{TRACE} model where the underlying continuous data $z_{ij}$ are generated as $z_i \sim t_{10}(\alpha^* + x_i^{\top} B^*, \Sigma^*)$, where  $ t_{10}(\mu, \Sigma)$ denotes a multivariate $t$ distribution with $10$ degrees of freedom with mean $\mu$ and scale matrix $\Sigma$ and have the same correlation structure as above, and
3) (common) data are simulated under the \texttt{TRACE} model where $\Sigma^* = \rho 1_p 1^{\top}_p + (1-\rho) I$ with $1_p$ a $p$-dimensional vector of all ones and $\rho$ drawn from a uniform distribution in $(0,0.8)$. The latter scenario includes cases where $\Sigma \approx I$, similar to the Indian buffet process model. The elements of the design matrix are simulated from $N(0,1)$ and the regression coefficients $\beta_{jl}^*$ are drawn from $N(0,1)$.
We also simulate synthetic data in the above three scenarios without covariates.

We compare the methods through estimation errors of $\pi_j^*$, $\Delta_{100|80}^*$, and $\Sigma^*$ using $\|\hat{\pi} - \pi^*\|/(np)$ and $\|\hat{\Sigma} - \Sigma^*\|/p^2$ for estimators $\hat{\pi}$ and $\hat{\Sigma}$, where $\|\cdot\|$ is the Frobenius norm. We use posterior means as estimators. In the setting without covariates, the error for the marginal occurrence probabilities is defined in terms of the vector $L_2$ norm.  In \texttt{IBP} type models with independent rows $E(p_n^*) = \sum_{j=1}^p w_j$ with $w_j \sim \textrm{Bernoulli}\{1-(1-\pi_j)^n\}$, implying $E(p^*_n) = \sum_{j=1}^p \{1-(1-\pi_j)^n\}$ from which the expected value of $\Delta_{m|n}^*$ can be calculated. In the regression setting, the dependence between species led to a more complex expression, so we obtain a Monte Carlo estimator of $\Delta_{m|n}^*$. We compute the mean square errors for each estimator $\hat{\Delta}_{100|80}$.

\begin{figure}
\begin{center}
\includegraphics[width=0.9\textwidth]{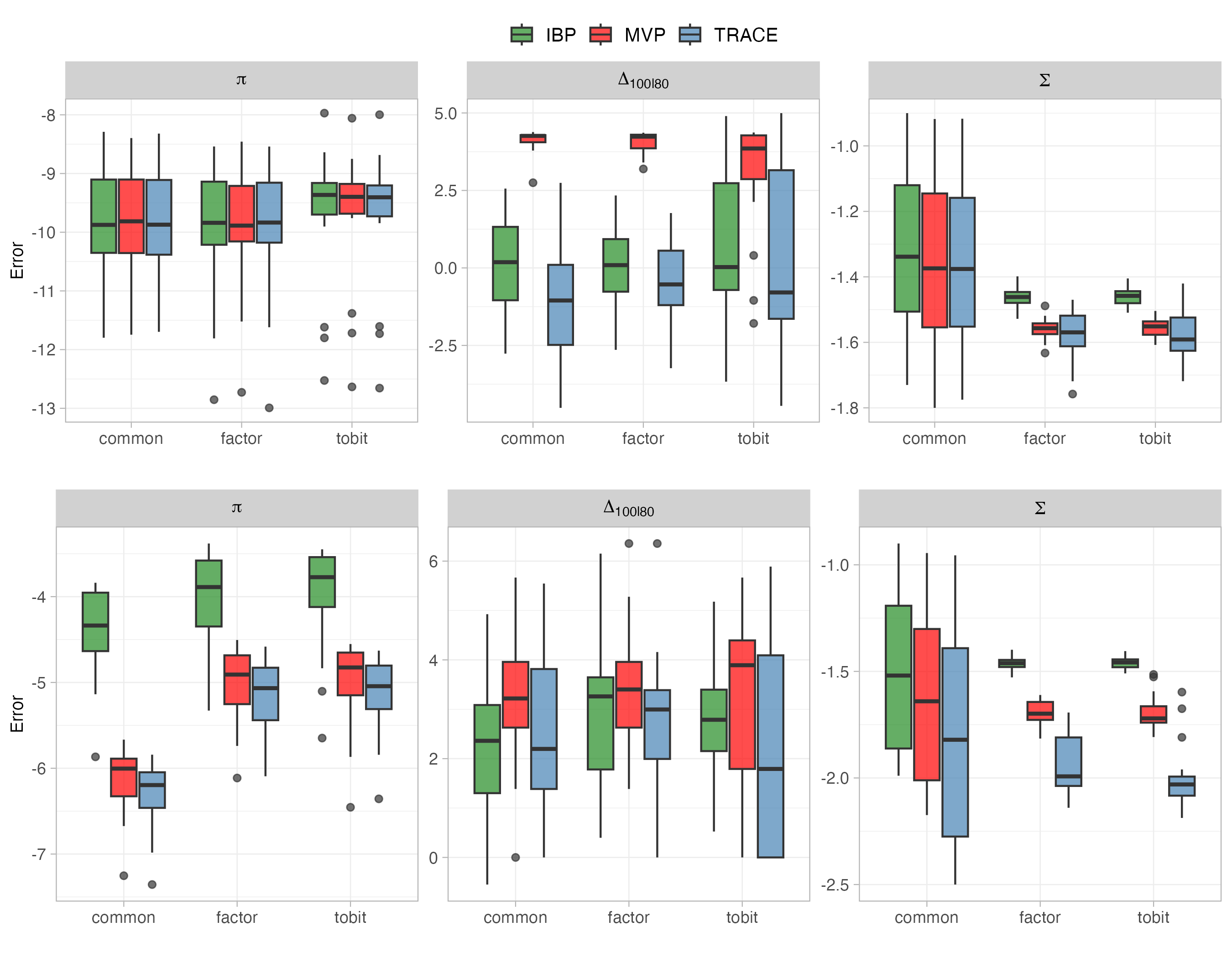}
\caption{Logarithm of Frobenius errors for $\pi^*$, $\Sigma^*$ and $\Delta_{100|80}^*$ of the \texttt{TRACE} (blue), \texttt{IBP} (green) and multivariate probit (red) models in 20 simulations for each scenario without (top panel) and with covariates (bottom panel).}
    \label{fig:sim_plot}
    \end{center}
\end{figure}

Figure \ref{fig:sim_plot} shows the results for the \texttt{TRACE}, \texttt{IBP} and multivariate probit models. The top and bottom panels show results without and with covariates, respectively. In all scenarios,  \texttt{TRACE} performs better or is comparable to the best competitor in all performance metrics. The multivariate probit model performs poorly in estimating $\Delta_{100|80}^*$. Since the \texttt{IBP} model assumes zero correlation and does not include covariates, it does poorly in estimating $\sigma_{jj'}$s and in inferring marginal occurrence probabilities in the regression setting. Figure \ref{fig:truncation} of the supplementary materials shows that the results for \texttt{TRACE} are robust to changes in the truncation level.

\section{Fungal biodiversity application}

We analyze data from a study on fungi biodiversity in Finland \citep{AbregoFungi}. Fungi are a highly diverse group of organisms, they play a major role in ecosystem functioning and are important for human health, food production, and conservation of nature.
The data consist of 134 samples from different sites in four study locations in Finland, with 29 samples in Site 1, 97 samples in Site 2, 2 samples in Site 3 and 6 samples in Site 4. 
A cyclone sampler was used to acquire spore samples from each study site.
Based on a probabilistic molecular species classifier using DNA meta-barcoding, $1021$ distinct species were identified. For each sample, the data include two covariates: a categorical variable with four levels indicating the study location and a continuous variable that provides information about the week.
To assess any temporal dependence or effect of habitat type, we include these covariates in the model as predictors. We encoded the site variable through three dummies with Site 2, which had the highest number of observations, as the reference.

Given the huge number of fungal species identified and the presence of many rare species, in implementing \texttt{TRACE}, we used the hierarchical shrinkage prior for the between species correlations described in Section \ref{sec:corr}. Similarly to the simulations, we center the distribution of $\gamma$ on the empirical mean of the number of species per sample. For the regression coefficients, we use a normal prior centered on zero, $\beta_j \sim N_q(0, \psi I)$. This choice led to a better fit to the data with respect to the hierarchical formulation for the regression coefficients defined in \eqref{eq:b_hier}, and is less computationally demanding. 

\begin{figure}[t]
\begin{center}
\includegraphics[width=0.9\textwidth]{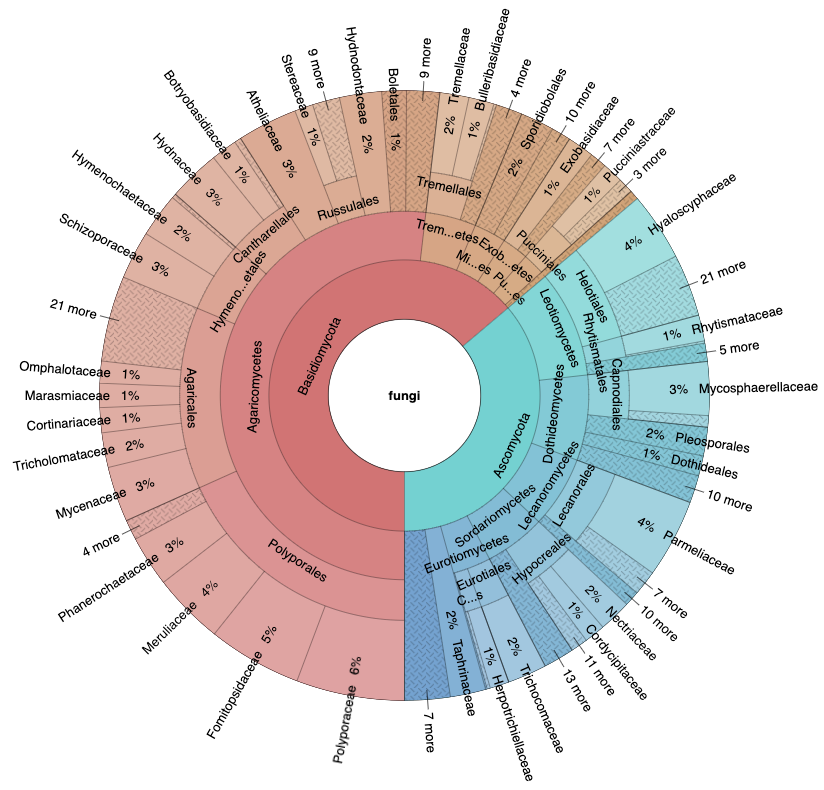}
\caption{Fungal taxonomic composition and marginal species occurrence probabilities.}
    \label{fig:krona}
    \end{center}
\end{figure}

We used Krona wheels \citep{ondov2011interactive} to show the taxonomic composition of the operational taxonomic units (\texttt{OTU}s) found and their relative abundance. As a measure of relative species abundance, we used the posterior mean of the marginal occurrence probabilities of each species. The results are shown in Figure \ref{fig:krona}.  Polyporales and Agaricales are the dominating orders of Basidiomycota, whereas Helotiales and Lecanorales are the dominating orders of Ascomycota. For a version of the Krona wheel showcasing changes as covariate values vary, see supplementary materials. It is interesting to see how these marginal occurrence probabilities change with location-specific covariates. For example, the proportion of Ascomycota and Basidiomycota remains similar in three sites, but differs at Site 4, where there are fewer Ascomycota species. This distinction may be related to the specific characteristics of Site 4, a small and highly isolated island without trees, separating it from the other sites, which are characterized by expansive natural spruce-dominated forests \citep{AbregoFungi}. The four locations in Finland exhibit significant differences in sample-specific species richness, as illustrated in Figure \ref{fig:spSite} of the supplementary material. Specifically, Site 1 shows a greater species richness than other sites, consistent with \cite{AbregoFungi}.

The correlation coefficients $\sigma_{jj'}$ capture the dependence in co-occurrence, which is of interest in studying species interactions. We flag species pairs that have 90\% credible intervals for $\sigma_{jj'}$ that do not include zero. Although it is difficult to interpret the results on the species level due to the large dimensionality, we can obtain interesting conclusions by grouping species that fall in the same branch of the taxonomic tree; for example, species of the same order. As an illustration, we focus on order pairs that exhibit the highest relative frequency of correlations different from zero. Relative frequency refers to the total positive (or negative) correlations between two orders among all possible pairs.
We notice that different orders of lichenized fungi \citep{lucking20172016} exhibit a positive correlation. In terms of positive relative frequency, the $33\%$ of (Lecideales, Pertusariales) and (Umbilicariales, Phaeomoniellales), the $25\%$ of (Lecideales, Baeomycetales) and the $23\%$ of (Xylariales, Lecideales) are correlated in occurrence. Perhaps, the environmental conditions that support one type of lichen are also favorable for other types. Another important group is mycorrhizal fungi, which are in symbiotic relationships with plant roots.
We notice a positive correlation ($50\%$ in terms of positive relative frequency) between the Sebacinales and Cantharellales orders of mycorrhizal fungi, probably due to the fact that both orders tend to form relationships with similar plant species such as orchids \citep{qin2020similar} which are widespread in Finland.

In biodiversity research, a key focus is inference on species richness, corresponding to the number of species in a community \citep{colwell2009biodiversity}. 
Identifying locations having unusually low or high biodiversity, measured in terms of species richness, is of key interest for conservationists. In addition, inferring locations that contain many undiscovered species is important to optimize future sampling efforts. For Sites 1 and 2 (the other sites have few samples), we randomly select test sets that contain $25\%$ of the observations. We fit each of the methods to the training data and use posterior predictive distributions to estimate the new species in the site-specific test sets. 
 We compare the results with those of the same competitors in Section \ref{sec:simulation}, while also fitting a three-parameter Indian buffet process \citep{teh2009indian} by maximum likelihood. We expect substantial heterogeneity between sites, so we fitted the two \texttt{IBP} models separately to the data from each site; pooled analyses led to worse predictive performance. For each model, site, and partial test data sample size, we calculated the mean square predictive error for the number of new species discovered $p_n^*$. Table \ref{tab:pnstar_MSE} shows the median and the interquartile range of these mean square errors. The \texttt{TRACE} model substantially outperforms competitors.

\begin{table}[t]
    \centering
    \caption{Performance of the \texttt{TRACE}, multivariate probit, Indian buffet process and three parameter Indian buffet process models in predicting discovery of new species of fungi in Finland for Site 1 and Site 2. The mean square error (MSE) is in logarithm scale.}
    \vspace{0.3cm}
    \begin{tabular}{lcccccccc}
    
   \multicolumn{1}{}{} &
      \multicolumn{4}{c}{Site 1} &
      \multicolumn{4}{c}{Site 2}  \\
      
     & TRACE & IBP &MVP& 3IBP & TRACE & IBP  &MVP & 3IBP\\
   MSE $p^*_{n}$ (median) & 4.75 &  8.78 &  7.86 &  7.02 & 5.81 & 8.19 & 7.50&6.90\\
MSE $p^*_{n}$ (IQR) & 0.09 & 0.04 & 0.14 &0.67 & 0.99  & 0.72 & 0.61 & 0.15\\
  \end{tabular}
    \label{tab:pnstar_MSE}
\end{table}

\section{Discussion}

In this article, we propose a novel class of models for inference on multivariate binary outcomes that have a growing dimension. Although the developments were concretely motivated by biodiversity studies, the modeling framework has a much broader applicability. The \texttt{TRACE} class maintains many of the desirable properties of the popular \texttt{IBP}, while addressing important statistical complexities that arise in applications, such as dependence in species occurrence, covariate effects, and the need to borrow information in conducting inferences for rare species. 

In introducing a class of \texttt{IBP}-type models that allow a flexible dependence structure between species, we are faced with the challenge of how to effectively characterize an increasing-dimensional correlation matrix. 
In fixed dimensional settings, it is common to rely on latent factor models to reduce dimensionality in characterizing dependence in multivariate binary data. However, such approaches cannot be directly applied in the \texttt{IBP} setting, which inevitably involves extremely sparse binary data and requires a coherent framework for prediction. 
For addressing this problem, we propose a hierarchical approach which effectively induces shrinkage in the modeling of the correlation matrix. This initial direction should motivate more work in this area; for example, alternative factor modeling approaches (perhaps related to \cite{beraha2023normalised, mauri2025factor}) or stochastic block model formulations that infer latent communities of species based on their correlation structure. 

There are many interesting directions for future research motivated directly by biodiversity studies. Firstly, most current \texttt{IBP} models impose a restrictive rate of discovery of new species as additional samples are taken. However, in exploratory data analyses, we found that fitting the empirical species accumulation curves from biodiversity studies often requires more flexibility. One promising direction in this regard is to introduce dependence in more sophisticated versions of the \texttt{IBP}, such as the three-parameter generalization with power-law behaviour of \cite{teh2009indian}. 

Another important direction is making our statistical tools computationally efficient and accessible so that ecologists can easily use them. Joint species distribution models have become routinely used in ecology, but their inference methods cannot handle new species discovery and/or the presence of many rare species in the sample. It is of substantial importance to be able to predict future species discovery based on current samples, while accurately characterizing uncertainty and accommodating effects of climate change and other covariates. It is also of substantial interest to infer which species are endangered and should be added to so-called red lists. Our statistical tools provide a first step towards automating such inferences and decisions.

\section*{Acknowledgement}
This project has received funding from the European Research Council under the European Union's Horizon 2020 research and innovation programme (grant agreement No 856506). The authors thank Filippo Ascolani for his helpful comments on the first version of this work.

\bibliography{reference}

\begin{thebibliography}{}

\bibitem[\protect\citeauthoryear{Abramowitz and Stegun}{Abramowitz and
  Stegun}{1948}]{AbraSteg72}
Abramowitz, M. and I.~A. Stegun (1948).
\newblock {\em Handbook of Mathematical Functions with Formulas, Graphs, and
  Mathematical Tables}, Volume~55.
\newblock US Government Printing Office.

\bibitem[\protect\citeauthoryear{Abrego, Furneaux, Hardwick, Somervuo,
  Palorinne, Aguilar-Trigueros, Andrew, Babiy, Bao, Bazzano, et~al.}{Abrego
  et~al.}{2024}]{abrego2024airborne}
Abrego, N., B.~Furneaux, B.~Hardwick, P.~Somervuo, I.~Palorinne, C.~A.
  Aguilar-Trigueros, N.~R. Andrew, U.~V. Babiy, T.~Bao, G.~Bazzano, et~al.
  (2024).
\newblock Airborne {DNA} reveals predictable spatial and seasonal dynamics of
  fungi.
\newblock {\em Nature\/}~{\em 631}, 835--842.

\bibitem[\protect\citeauthoryear{Abrego, Norros, Halme, Somervuo, Ali-Kovero,
  and Ovaskainen}{Abrego et~al.}{2018}]{AbregoFungi}
Abrego, N., V.~Norros, P.~Halme, P.~Somervuo, H.~Ali-Kovero, and O.~Ovaskainen
  (2018).
\newblock Give me a sample of air and {I} will tell which species are found
  from your region: Molecular identification of fungi from airborne spore
  samples.
\newblock {\em Mol. Ecol. Res.\/}~{\em 18\/}(3), 511--524.

\bibitem[\protect\citeauthoryear{Beraha and Griffin}{Beraha and
  Griffin}{2023}]{beraha2023normalised}
Beraha, M. and J.~E. Griffin (2023).
\newblock Normalised latent measure factor models.
\newblock {\em J. R. Statist. Soc. B\/}~{\em 85\/}(4), 1247--1270.

\bibitem[\protect\citeauthoryear{Broderick, Pitman, and Jordan}{Broderick
  et~al.}{2013}]{Broderick2013FeatureAP}
Broderick, T., J.~Pitman, and M.~I. Jordan (2013).
\newblock Feature allocations, probability functions, and paintboxes.
\newblock {\em Bayesian Anal.\/}~{\em 8}, 801--836.

\bibitem[\protect\citeauthoryear{Camerlenghi, Favaro, Masoero, and
  Broderick}{Camerlenghi et~al.}{2024}]{camerlenghi2022scaled}
Camerlenghi, F., S.~Favaro, L.~Masoero, and T.~Broderick (2024).
\newblock Scaled process priors for {B}ayesian nonparametric estimation of the
  unseen genetic variation.
\newblock {\em J. Am. Statist. Assoc.\/}~{\em 119\/}(545), 320--331.

\bibitem[\protect\citeauthoryear{Chakraborty, Ou, and Dunson}{Chakraborty
  et~al.}{2024}]{chakraborty2022bayesian}
Chakraborty, A., R.~Ou, and D.~B. Dunson (2024).
\newblock Bayesian inference on high-dimensional multivariate binary responses.
\newblock {\em J. Am. Statist. Assoc.\/}~{\em 119\/}(548), 2560--2571.

\bibitem[\protect\citeauthoryear{Chib and Greenberg}{Chib and
  Greenberg}{1998}]{ChibMultivariateProbit}
Chib, S. and E.~Greenberg (1998).
\newblock {Analysis of multivariate probit models}.
\newblock {\em Biometrika\/}~{\em 85\/}(2), 347--361.

\bibitem[\protect\citeauthoryear{Colwell}{Colwell}{2009}]{colwell2009biodiversity}
Colwell, R.~K. (2009).
\newblock Biodiversity: concepts, patterns, and measurement.
\newblock {\em The Princeton Guide to Ecology\/}~{\em 663}, 257--263.

\bibitem[\protect\citeauthoryear{Compson, McClenaghan, Singer, Fahner, and
  Hajibabaei}{Compson et~al.}{2020}]{compson2020metabarcoding}
Compson, Z.~G., B.~McClenaghan, G.~A. Singer, N.~A. Fahner, and M.~Hajibabaei
  (2020).
\newblock Metabarcoding from microbes to mammals: comprehensive bioassessment
  on a global scale.
\newblock {\em Front. Ecol. Evol.\/}~{\em 8}, 581835.

\bibitem[\protect\citeauthoryear{Di~Benedetto, Caron, and Teh}{Di~Benedetto
  et~al.}{2020}]{di2020non}
Di~Benedetto, G., F.~Caron, and Y.~W. Teh (2020).
\newblock Non-exchangeable feature allocation models with sublinear growth of
  the feature sizes.
\newblock In {\em Proc. Int. Conf. Artif. Intel. Statist.}, Volume 108, pp.\
  3208--3218. PMLR.

\bibitem[\protect\citeauthoryear{Doshi, Miller, Van~Gael, and Teh}{Doshi
  et~al.}{2009}]{doshi2009variational}
Doshi, F., K.~Miller, J.~Van~Gael, and Y.~W. Teh (2009).
\newblock Variational inference for the {I}ndian buffet process.
\newblock In {\em Proc. 12th Int. Conf. Artif. Intel. Statist.}, Volume~5, pp.\
   137--144. PMLR.

\bibitem[\protect\citeauthoryear{Drezner and Wesolowsky}{Drezner and
  Wesolowsky}{1990}]{drezner1990computation}
Drezner, Z. and G.~O. Wesolowsky (1990).
\newblock On the computation of the bivariate normal integral.
\newblock {\em J. Statist. Comput. Simul.\/}~{\em 35\/}(1-2), 101--107.

\bibitem[\protect\citeauthoryear{Gershman, Frazier, and Blei}{Gershman
  et~al.}{2014}]{gershman2014distance}
Gershman, S.~J., P.~I. Frazier, and D.~M. Blei (2014).
\newblock Distance dependent infinite latent feature models.
\newblock {\em IEEE Trans. Neural Networks Learn. Syst.\/}~{\em 37\/}(2),
  334--345.

\bibitem[\protect\citeauthoryear{Griffiths and Ghahramani}{Griffiths and
  Ghahramani}{2011}]{JMLR:v12:griffiths11a}
Griffiths, T. and Z.~Ghahramani (2011).
\newblock The {I}ndian buffet process: An introduction and review.
\newblock {\em J. Mach. Learn. Res.\/}~{\em 12\/}(32), 1185--1224.

\bibitem[\protect\citeauthoryear{Ishwaran and James}{Ishwaran and
  James}{2001}]{ishwaran2001gibbs}
Ishwaran, H. and L.~F. James (2001).
\newblock Gibbs sampling methods for stick-breaking priors.
\newblock {\em J. Am. Statist. Assoc.\/}~{\em 96\/}(453), 161--173.

\bibitem[\protect\citeauthoryear{Johndrow, Smith, Pillai, and Dunson}{Johndrow
  et~al.}{2019}]{JohndrowImbalanced}
Johndrow, J.~E., A.~Smith, N.~Pillai, and D.~B. Dunson (2019).
\newblock Mcmc for imbalanced categorical data.
\newblock {\em J. Am. Statist. Assoc.\/}~{\em 114\/}(527), 1394--1403.

\bibitem[\protect\citeauthoryear{Lewandowski, Kurowicka, and Joe}{Lewandowski
  et~al.}{2009}]{lewandowski2009generating}
Lewandowski, D., D.~Kurowicka, and H.~Joe (2009).
\newblock Generating random correlation matrices based on vines and extended
  onion method.
\newblock {\em J. Multiv. Anal.\/}~{\em 100\/}(9), 1989--2001.

\bibitem[\protect\citeauthoryear{L{\"u}cking, Hodkinson, and
  Leavitt}{L{\"u}cking et~al.}{2017}]{lucking20172016}
L{\"u}cking, R., B.~P. Hodkinson, and S.~D. Leavitt (2017).
\newblock The 2016 classification of lichenized fungi in the {A}scomycota and
  {B}asidiomycota - {A}pproaching one thousand genera.
\newblock {\em The Bryologist\/}~{\em 119\/}(4), 361--416.

\bibitem[\protect\citeauthoryear{Mauri and Dunson}{Mauri and
  Dunson}{2025}]{mauri2025factor}
Mauri, L. and D.~B. Dunson (2025).
\newblock Factor pre-training in {B}ayesian multivariate logistic models.
\newblock {\em arXiv:2409.17441\/}.

\bibitem[\protect\citeauthoryear{Ondov, Bergman, and Phillippy}{Ondov
  et~al.}{2011}]{ondov2011interactive}
Ondov, B.~D., N.~H. Bergman, and A.~M. Phillippy (2011).
\newblock Interactive metagenomic visualization in a {W}eb browser.
\newblock {\em BMC Bioinform.\/}~{\em 12\/}(385), 1--10.

\bibitem[\protect\citeauthoryear{Ovaskainen, Abrego, Halme, and
  Dunson}{Ovaskainen et~al.}{2016}]{OvaskainenEtal2016}
Ovaskainen, O., N.~Abrego, P.~Halme, and D.~B. Dunson (2016).
\newblock Using latent variable models to identify large networks of
  species-to-species associations at different spatial scales.
\newblock {\em Meth. Ecol. Evol.\/}~{\em 7\/}(5), 549--555.

\bibitem[\protect\citeauthoryear{Owen}{Owen}{1980}]{owen1980table}
Owen, D.~B. (1980).
\newblock A table of normal integrals.
\newblock {\em Communic. Statist. Simul. Comput.\/}~{\em 9\/}(4), 389--419.

\bibitem[\protect\citeauthoryear{Papoulis and Pillai}{Papoulis and
  Pillai}{2002}]{papoulis2002probability}
Papoulis, A. and S.~Pillai (2002).
\newblock {\em {Probability, Random Variables, and Stochastic Processes}}.
\newblock McGraw-Hill series in electrical and computer engineering.
  McGraw-Hill.

\bibitem[\protect\citeauthoryear{Qin, Zhang, Zhang, and Wang}{Qin
  et~al.}{2020}]{qin2020similar}
Qin, J., W.~Zhang, S.-B. Zhang, and J.-H. Wang (2020).
\newblock Similar mycorrhizal fungal communities associated with epiphytic and
  lithophytic orchids of {C}oelogyne corymbosa.
\newblock {\em Plant Diversity\/}~{\em 42\/}(5), 362--369.

\bibitem[\protect\citeauthoryear{Somervuo, Koskela, Pennanen, Henrik~Nilsson,
  and Ovaskainen}{Somervuo et~al.}{2016}]{somervuo2016unbiased}
Somervuo, P., S.~Koskela, J.~Pennanen, R.~Henrik~Nilsson, and O.~Ovaskainen
  (2016).
\newblock Unbiased probabilistic taxonomic classification for {DNA} barcoding.
\newblock {\em Bioinformatics\/}~{\em 32\/}(19), 2920--2927.

\bibitem[\protect\citeauthoryear{Teh and Gorur}{Teh and
  Gorur}{2009}]{teh2009indian}
Teh, Y. and D.~Gorur (2009).
\newblock Indian buffet processes with power-law behavior.
\newblock In {\em Advanc. Neur. Inform. Process. Syst.}, Volume~22. Curran
  Associates, Inc.

\bibitem[\protect\citeauthoryear{Thibaux and Jordan}{Thibaux and
  Jordan}{2007}]{thibaux2007hierarchical}
Thibaux, R. and M.~I. Jordan (2007).
\newblock Hierarchical beta processes and the {I}ndian buffet process.
\newblock In {\em Proc. 11th Int. Conf. Artif. Intel. Statist.}, Volume~2, pp.\
   564--571. PMLR.

\bibitem[\protect\citeauthoryear{Tierney and Kadane}{Tierney and
  Kadane}{1986}]{tierney1986accurate}
Tierney, L. and J.~B. Kadane (1986).
\newblock Accurate approximations for posterior moments and marginal densities.
\newblock {\em J. Am. Statist. Assoc.\/}~{\em 81\/}(393), 82--86.

\bibitem[\protect\citeauthoryear{Tikhonov, Opedal, Abrego, Lehikoinen,
  de~Jonge, Oksanen, and Ovaskainen}{Tikhonov et~al.}{2020}]{TikhonovHmsc20200}
Tikhonov, G., O.~H. Opedal, N.~Abrego, A.~Lehikoinen, M.~M.~J. de~Jonge,
  J.~Oksanen, and O.~Ovaskainen (2020).
\newblock Joint species distribution modelling with the r-package {HMSC}.
\newblock {\em Meth. Ecol. Evol.\/}~{\em 11\/}(3), 442--447.

\bibitem[\protect\citeauthoryear{Warr, Dahl, Meyer, and Lui}{Warr
  et~al.}{2022}]{warr2022attraction}
Warr, R.~L., D.~B. Dahl, J.~M. Meyer, and A.~Lui (2022).
\newblock The attraction {I}ndian buffet distribution.
\newblock {\em Bayesian Anal.\/}~{\em 17\/}(3), 931--967.

\bibitem[\protect\citeauthoryear{Williamson, Orbanz, and Ghahramani}{Williamson
  et~al.}{2010}]{williamson2010dependent}
Williamson, S., P.~Orbanz, and Z.~Ghahramani (2010).
\newblock Dependent {I}ndian buffet processes.
\newblock In {\em Proc. 13th Int. Conf. Artif. Intel. Statist.}, Volume~9, pp.\
   924--931. PMLR.

\end{thebibliography}

\newpage
\section*{Supplementary materials}

\setcounter{section}{0}
\setcounter{table}{0}
\setcounter{figure}{0}
\setcounter{equation}{0}
\renewcommand{\thefigure}{S\arabic{figure}}
\renewcommand{\thesection}{S.\arabic{section}}
\renewcommand{\thetable}{S\arabic{table}}
\renewcommand{\theequation}{S.\arabic{equation}}

\section{Proofs}

\subsection{Proof of Theorem 1} \label{sec:proof_ni}

We start by computing the $p\to \infty$ limit of the expected value of $n_{i+} = \sum_{j=1}^p y_{ij}$ for the no covariates \texttt{TRACE} model. Let $\phi(\cdot)$ denote the density function of a standard normal random variable. The expected value of the number of species per sample $n_{i+}$ for each $i=1,\dots,n$ is 

\begin{equation} \label{eq:exp_ni}
    E(n_{i+}) = 
     \sum_{j=1}^p \int_{-\infty}^{\infty} \Phi(\alpha_j) \frac{1}{\tau_p}\phi\bigg(\frac{\alpha_j - \mu_p}{\tau_p}\bigg) \,d\alpha_j 
     =  \sum_{j=1}^p \int_{-\infty}^{\infty} \Phi(\tau_p \alpha_j + \mu_p) \phi(\alpha_j) \,d\alpha_j.
\end{equation}
\cite{owen1980table} showed that \eqref{eq:exp_ni} is equal to $ \sum_{j=1}^p \Phi \{\mu_p/(1+\tau_p^2)^{1/2}\}$. Then, by  substituting $\mu_p$ in the latter expression we obtain
\begin{equation*}
E(n_{i+})  = p \frac{\gamma/p}{\gamma/p + 1},
\end{equation*}
which converges to $\gamma$ as $p \to \infty$ for each $i=1,\dots,n$.

If we further assume $\Sigma = I$, then $n_{i+}$ is a sum of independent Bernoulli random variables and thus has a binomial distribution with size $p$ and success probability $(\gamma/p)/(\gamma/p + 1)$. Given that $p\{(\gamma/p)/(\gamma/p + 1)\}$ converges to $\gamma$ for $p \to \infty$,  it follows by the law of rare events (see for instance page 113 of \cite{papoulis2002probability}) that $n_{i+}$ converges in distribution to a Poisson random variable with parameter $\gamma$.

\subsection{Proof of Proposition 1}

We start by proving the following lemma, which provides a useful result for the expected value of the product of $y_{ij}$ and $y_{ij'}$ with $j\neq j'$ when $p$ grows in the no covariates \texttt{TRACE} model.

\begin{lemma}
    For the no covariates \texttt{TRACE} model the following equation holds:
    \begin{equation*}
        \lim_{p\to\infty} \sum_{j<j'\le p}  E(y_{ij} y_{ij'}) = \frac{\gamma^2}{2} + \lim_{p\to\infty} \frac{\gamma^2}{(\gamma+p)^2}  
        \sum_{j<j'\le p}  |\exp(\sigma_{jj'})-1|.
    \end{equation*}
\end{lemma}

\begin{proof}
The expected value of the product of $y_{ij}$ and $y_{ij'}$, with $j\neq j'$, is equal to
\begin{equation} \label{eq:exp_prod}
     E(y_{ij} y_{ij'})  = \int_{-\infty}^{\infty} \int_{-\infty}^{\infty} \Phi_2(\alpha; \sigma_{jj'})  \frac{1}{\tau_p^2} \phi\bigg(\frac{\alpha_j - \mu_p}{\tau_p}\bigg)  \phi\left(\frac{\alpha_{j'} - \mu_p}{\tau_p}\right) \,d\alpha_j \,d\alpha_{j'},
\end{equation}
 where $\Phi_2(\alpha; \sigma_{jj'})$ is the cumulative distribution function of a bivariate normal evaluated at $\alpha = (\alpha_j, \alpha_{j'})^{\top}$ with mean zero and correlation $\sigma_{jj'}$. 
 \cite{drezner1990computation} showed that 
the cumulative distribution function of a bivariate normal can be written as
\begin{equation*}
    \Phi_2(\alpha; \sigma_{jj'}) = \Phi(\alpha_j)\Phi(\alpha_{j'}) + \frac{1}{2\pi} I(\sigma_{jj'}),
\end{equation*}
\begin{equation*}
        I(\sigma_{jj'})= \int_{0}^{\sigma_{jj'}} \frac{1}{(1-u^2)^{1/2}}\exp\bigg\{-\frac{1}{2(1-u^2)}(\alpha_j^2-2\alpha_j \alpha_{j'} u + \alpha_{j'}^2)\bigg\} du.
\end{equation*}
Then, substituting the above expression for the bivariate cumulative distribution function and leveraging the same result on integrals of normal cumulative distribution functions \citep{owen1980table} used in the proof of Theorem 1, equation \eqref{eq:exp_prod} becomes
\begin{equation*}
    E(y_{ij} y_{ij'}) = \frac{\gamma^2}{(\gamma+p)^2} + \frac{1}{2\pi} \int_{-\infty}^{\infty} \int_{-\infty}^{\infty} I(\sigma_{jj'}) \frac{1}{\tau_p} \phi\left(\frac{\alpha_j - \mu_p}{\tau_p}\right)  \frac{1}{\tau_p}\phi\left(\frac{\alpha_{j'} - \mu_p}{\tau_p}\right) \,d\alpha_j \,d\alpha_{j'}.
\end{equation*}
The latter result implies that, as $p$ grows, the quantity $\lim_{p\to\infty} \sum_{j<j'\le p}  E(y_{ij} y_{ij'})$ is equal to 
\begin{equation} \label{eq:limp_vniSI}
\frac{\gamma^2}{2} + \lim_{p \to \infty} \sum_{j<j'\le p} \frac{1}{2\pi} \int_{-\infty}^{\infty} \int_{-\infty}^{\infty} I(\sigma_{jj'}) \frac{1}{\tau_p} \phi\left(\frac{\alpha_j - \mu_p}{\tau_p}\right)  \frac{1}{\tau_p}\phi\left(\frac{\alpha_{j'} - \mu_p}{\tau_p}\right) \,d\alpha_j \,d\alpha_{j'}.
\end{equation}
By the dominated convergence theorem, we first compute the limit of the integrand in $\eqref{eq:limp_vniSI}$ and then do the integration. 
Let $a_p \simeq b_p$ indicate that $\lim_{p \to \infty} a_p/b_p = 1$, that is $a_p$ and $b_p$ have the same order. Exploiting the following asymptotic expansion \citep{AbraSteg72} for the inverse of the cumulative distribution function of a normal distribution
\begin{equation*}
    \Phi^{-1}(x) = -\{-2\log x -\log(-2\log x) - \log(2\pi)\}^{1/2}+o(1), \quad x \rightarrow 0 ,
\end{equation*}
we have $\mu_p \simeq -[(1+\tau_p^2)\{-2\log \{\gamma/(\gamma+p)\} -\log[-2\log \{\gamma/(\gamma+p)\} ] - \log(2\pi)\}]^{-1/2}$ for $p \rightarrow \infty$. 
Then, by standard convergence theorems we obtain 
\begin{equation*}
     \frac{1}{\tau_p}  \phi\left(\frac{\alpha_j - \mu_p}{\tau_p}\right)  \frac{1}{\tau_p}  \phi\left(\frac{\alpha_j' - \mu_p}{\tau_p}\right) \simeq  \exp\{-(1+\alpha_j+\alpha_j')\}  \frac{\gamma^2}{(\gamma+p)^2},
\end{equation*}
from which follows that the limit in \eqref{eq:limp_vniSI} is equivalent to 
\begin{equation*}
      \frac{\gamma^2}{2} + \lim_{p \to \infty} \sum_{j<j'\le p} \exp(-1) \frac{\gamma^2}{(\gamma+p)^2}   \frac{1}{2\pi} \int_{-\infty}^{\infty} \int_{-\infty}^{\infty} I(\sigma_{jj'})\exp(-\alpha_j) \exp(-\alpha_j') \,d\alpha_j d\alpha_{j'}.
\end{equation*}
Interchanging the order of integration with the Fubini-Tonelli theorem, the integrals in the latter expression become
\begin{equation} \label{eq:int_triple}
    \int_{0}^{\sigma_{jj'}}  \int_{-\infty}^{\infty} \int_{-\infty}^{\infty} \frac{1}{(1-u^2)^{1/2}}\exp\bigg\{-\frac{\alpha_j^2-2\alpha_j \alpha_{j'} u + \alpha_{j'}^2}{2(1-u^2)}\bigg\} \exp(-\alpha_j-\alpha_{j'})  \,d\alpha_j \,d\alpha_{j'} du.
\end{equation}
By completing the square for both $\alpha_j$ and $\alpha_j'$, we recognized the kernels of two normal densities and the subsequent integration with respect to both $\alpha_j$ and $\alpha_j'$ becomes straightforward. Then, by further algebraic simplifications  \eqref{eq:int_triple} is equal to $2\pi \exp(1) \int_{0}^{\sigma_{jj'}} \exp(u) du = 2\pi \exp(1)|\exp(\sigma_{jj'})-1|$. We included the absolute value in the latter expression to consider both positive and negative correlations. Thus, as $p$ grows we have
 \begin{equation*}
        \lim_{p\to\infty} \sum_{j<j'\le p}  E(y_{ij} y_{ij'}) = \frac{\gamma^2}{2} + \lim_{p\to\infty} \frac{\gamma^2}{(\gamma+p)^2}  
        \sum_{j<j'\le p}  |\exp(\sigma_{jj'})-1|,
    \end{equation*}
which concludes the proof.
\end{proof}

We now compute the variance of $n_{i+}$ as $p \to \infty$ for the no covariates \texttt{TRACE} model exploiting Lemma 1. 
Recalling that the observations for different species are not independent, the target object to compute is
\begin{equation}\label{eq:SIgoal_vni}
    \lim_{p\to\infty} \V(n_{i+}) = \lim_{p\to\infty} \sum_{j=1}^p\V(y_{ij}) + \lim_{p\to\infty} \sum_{j<j'\le p} \cov(y_{ij}, y_{ij'}).
\end{equation}
Applying the law of total variance, the variance of the observations is
\begin{equation*}
     \V(y_{ij})   =E[\Phi(\alpha_j)\{1-\Phi(\alpha_j)\}] + \V\{\Phi(\alpha_j)\} = E\{\Phi(\alpha_j)\} - E\{\Phi(\alpha_j)\}^2 = \frac{\gamma}{\gamma + p}-\frac{\gamma^2}{(\gamma + p)^2},
\end{equation*}
from which it follows that $\lim_{p \to \infty} \sum_{j=1}^p \V(y_{ij}) = \gamma$.  This is the variance of $n_{i+}$ if one assumes independence among species, so the covariance between $y_{ij}$ and $y_{ij'}$ is zero. Hence, the target equation in \eqref{eq:SIgoal_vni} is equivalent to 
\begin{equation*}
    \lim_{p\to\infty} \V(n_{i+}) = \gamma + \lim_{p\to\infty} \sum_{j<j'\le p} \cov(y_{ij}, y_{ij'}),
\end{equation*}
where
\begin{equation*}
    \cov(y_{ij}, y_{ij'}) = E(y_{ij} y_{ij'}) - E\{\Phi(\alpha_j)\}^2 = E(y_{ij} y_{ij'}) - \frac{\gamma^2}{(\gamma + p)^2}.
\end{equation*}
By computing the sum of the limits of the latter equation as $p\to \infty$, we obtain the following expression for the limit of the variance of $n_{i+}$ as $p$ grows
\begin{align*}
    \lim_{p\to\infty} \V(n_{i+}) &= \gamma  -\frac{\gamma^2}{2} +\lim_{p\to\infty} \sum_{j<j'\le p}  E(y_{ij} y_{ij'})\\
    & = \gamma + \lim_{p\to\infty} \frac{\gamma^2}{(\gamma+p)^2}  
        \sum_{j<j'\le p}  |\exp(\sigma_{jj'})-1|,
\end{align*}
where the latter equation follows from Lemma 1.
If we assume a finite number of correlation coefficients, $\sigma_{jj'} \in \{\rho_1,\ldots,\rho_K\}$ with $K \in \mathds{N}$, then $\sum_{j<j'\le p}  |\exp(\sigma_{jj'})-1| = \sum_{k=1}^K w_k |\exp(\rho_k)-1|$, with $w_k = \sum_{j<j'\le p} \mathds{1}(\sigma_{jj'}=\rho_k)$ and $\sum_{k=1}^K w_k = p(p-1)/2$.  
Considering that $\sum_{k=1}^K w_k |\exp(\rho_k)-1|$ is bounded above by $\{\underset{k}{\mathrm{max}} \, |\exp(\rho_k)-1|\} p(p-1)/2$ and below by $\{\underset{k}{\mathrm{min}} \, |\exp(\rho_k)-1|\} p(p-1)/2$, we conclude that
\begin{equation*}
    \gamma + \gamma^2/2 \, \{\underset{k}{\mathrm{min}}\, |\exp(\rho_k) -1|\} \le \lim_{p \to \infty} \V(n_{i+}) \le \gamma + \gamma^2/2 \, \{\underset{k}{\mathrm{max}} \, |\exp(\rho_k)-1|\}.
\end{equation*}

\subsection{Proof of Theorem 2}

The expected value of the number of species per site $i$, $n_{i+} = \sum_{j=1}^p y_{ij}$, for the \texttt{TRACE} model is 

\begin{equation} \label{eq:exp_nicov}
    E(n_{i+}) =  \sum_{j=1}^p E(y_{ij}) = 
     \sum_{j=1}^p \int_{-\infty}^{\infty} \Phi(u_{ij}) \frac{1}{(\tau_p^2 + x_i^{\top}\Psi x_i)^{1/2}}\phi\bigg\{\frac{u_{ij} - x_i^{\top}\nu- \mu_p}{(\tau_p^2 + x_i^{\top}\Psi x_i)^{1/2}}\bigg\} \,du_{ij}, 
\end{equation}
with $u_{ij} = \alpha_j + x_i^{\top}\beta_j$. \cite{owen1980table} showed that \eqref{eq:exp_nicov} is equal to 
\begin{equation*} 
 \sum_{j=1}^p \Phi \bigg\{ \frac{\mu_p + x_i^{\top}\nu}{(1+\tau_p^2 + x_i^{\top}\Psi x_i)^{1/2}}\bigg\}.  
\end{equation*}
The $p\to \infty$ limit of the latter equation can be computed by leveraging  the following asymptotic expansion \citep{AbraSteg72} for the cumulative distribution function of a normal distribution: $1-\Phi(x) = \{\exp(-x^2/2)\}/\{(2\pi)^{1/2}x\} + o(1/x^2)$, for $x \rightarrow \infty$. 
Therefore, as $p$ grows the expected value of $n_{i+}$ is
\begin{equation} \label{eq:lim_nireg}
    \lim_{p \rightarrow \infty} E(n_{i+}) = \lim_{p \rightarrow \infty} \frac{p (\tau_p^2+1+ x_i^{\top}\Psi x_i)^{1/2}}{(2\pi)^{1/2}(-\mu_p-x_i^{\top}\nu)} \exp\bigg\{-\frac{(\mu_p+x_i^{\top}\nu)^2}{2(\tau_p^2+1+ x_i^{\top}\Psi x_i)}\bigg\}.
\end{equation}
Employing the same asymptotic expansions for $\mu_p$ used in the proof of Proposition 1, the quantity  $\exp[-(\mu_p+x_i^{\top}\nu)^2/ \{2(\tau_p^2+1+ x_i^{\top}\Psi x_i)^{1/2}\} ]$ as $p \to \infty$ is equivalent to  
\begin{equation*}
    \exp \bigg\{\frac{2\log(p)}{2 \log(p) + x_i^{\top}\Psi x_i + 1} \log \bigg(\frac{\gamma}{\gamma + p} \bigg) \bigg\} 
   \bigg\{-2\log \bigg(\frac{\gamma}{\gamma+p}\bigg)\bigg\}^{1/2} (2\pi)^{1/2}\,\exp(x_i^{\top}\nu-1/2),
\end{equation*}
and analogously the quantity $(\tau_p^2+1+ x_i^{\top}\Psi x_i)^{1/2}/(-\mu_p-x_i^{\top}\nu)$ for $p\to \infty$ is equivalent to
\begin{equation*}
    \frac{\{ 1 + 2\log(p) + x_i^{\top}\Psi x_i \}^{1/2}}{2\log(p) - x_i^{\top}\nu}.
\end{equation*}
By further observing that by standard convergence theorems 
\begin{equation*}
    \lim_{p\to\infty} \left\{-2\log \bigg(\frac{\gamma}{\gamma+p}\right)\bigg\}^{1/2}     \frac{\{ 1 + 2\log(p) + x_i^{\top}\Psi x_i \}^{1/2}}{2\log(p) - x_i^{\top}\nu} = 1,
\end{equation*}
it follows that the limit in \eqref{eq:lim_nireg} by applying the product of limits is equal to
\begin{align*}
\lim_{p \rightarrow \infty} E(n_{i+}) & =
    \lim_{p\to\infty} p \exp(x_i^{\top}\nu-1/2) \exp \bigg\{\frac{2\log(p)}{2 \log(p) + x_i^{\top}\Psi x_i + 1} \log \bigg(\frac{\gamma}{\gamma + p} \bigg) \bigg\}\\
    &= \gamma \exp\{1/2+0.5(x_i^{\top}\Psi x_i)\} \exp(x_i^{\top}\nu-1/2) \\
    & = \gamma \exp\{x_i^{\top}\nu + 0.5(x_i^{\top}\Psi x_i)\},
\end{align*}
which concludes the proof of the first point of Theorem 2. 

The second point of Theorem 2 refers to the limiting variance of $n_{i+} = \sum_{j=1}^p y_{ij}$, which is 
\begin{equation*}
    \lim_{p\to\infty} \V(n_{i+}) = \lim_{p\to\infty} \sum_{j=1}^p\V(y_{ij}) + \lim_{p\to\infty} \sum_{j<j'\le p} \cov(y_{ij}, y_{ij'}).
\end{equation*}
By observing that $\lim_{p \to \infty}p E\{\Phi(\alpha_j + x_i^{\top}\beta_j)\} = \gamma g_i $ and 
$\lim_{p \to \infty} \{ p(p-1)/2 \} E\{\Phi(\alpha_j + x_i^{\top}\beta_j)\} = (\gamma g_i)^2/2$, 
with 
$g_i = \exp\{ x_i^{\top}\nu + (1/2) x_i^{\top}\Psi x_i\}$, the above limit is straightforward to compute adapting the same steps followed in the proof of Proposition 1.

\subsection{Proof of Proposition 2}

By observing that $ \pr(y_{ij}=1) = \int_{-\infty}^{\infty} \Phi(u_{ij})/ (\tau_p^2 + x_i^{\top}\Psi x_i)^{1/2}\phi\{(u_{ij} - x_i^{\top}\nu- \mu_p)/(\tau_p^2 + x_i^{\top}\Psi x_i)^{1/2}\} \,du_{ij}$,   the result follows immediately from the proof of Theorem 2.

\subsection{Proof of Proposition 3}

The target is computing the $p\to \infty$ limit of the expected value of the number of common species for the no covariates \texttt{TRACE} model
\begin{equation*}
    E(\commons) =  \sum_{j=1}^p \pr\{\Phi(\alpha_j) > \epsilon\} = \sum_{j=1}^p \pr\{\alpha_j > \Phi^{-1}(\epsilon)\} = p \bigg[1- \Phi \bigg\{\frac{\Phi^{-1}(\epsilon) - \mu_p}{\tau_p}\bigg\}\bigg].
\end{equation*}
Exploiting the same asymptotic expansion for the cumulative distribution function of a normal distribution used in the proof of Theorem 2, 
the limit of the expected  value of $\commons$ is
\begin{equation} \label{eq:lim_comm}
    \lim_{p \rightarrow \infty} E(\commons) = \lim_{p \rightarrow \infty} \frac{p \tau_p}{(2\pi)^{1/2}(\epsilon^*-\mu_p)} \exp\bigg\{-\frac{1}{2\tau_p^2}(\epsilon^* - \mu_p)^2\bigg\},
\end{equation}
with $\epsilon^* = \Phi^{-1}(\epsilon)$.
Then, by leveraging the same asymptotic expansion for the inverse of the
cumulative distribution function of a normal distribution used in the proof of Proposition 1 for $\mu_p$, as $p \to \infty$ we have
\begin{equation*}
    \exp\bigg\{-\frac{1}{2\tau_p^2}(\epsilon^* - \mu_p)^2\bigg\} \simeq \exp(-\epsilon^*)  \bigg(\frac{\gamma}{\gamma+p}\bigg) \bigg\{-2\log \bigg(\frac{\gamma}{\gamma+p}\bigg)\bigg\}^{1/2} (2\pi)^{1/2}\,\exp(-1/2),
\end{equation*}
\begin{equation*}
    \frac{\tau_p}{\epsilon^* - \mu_p} \simeq \frac{1}{[-2\log \{\gamma/(\gamma+p)\}]^{1/2}},
\end{equation*}
from which follows that the limit in \eqref{eq:lim_comm} is equal to 
\begin{equation*}
    \lim_{p \rightarrow \infty} E(\commons) = \lim_{p \rightarrow \infty} \exp\bigg(-\epsilon^*-\frac{1}{2}\bigg)p \bigg(\frac{\gamma}{\gamma+p}\bigg) = \gamma \exp\left(-\epsilon^*-\frac{1}{2}\right).
\end{equation*}

\subsection{Proof of Proposition 4}
The expected value of the number of common species for sample $i$ for the \texttt{TRACE} model is
\begin{equation*}
    E(c_{i}^{\epsilon}) =  \sum_{j=1}^p \pr\{\Phi(\alpha_j + x_i^{\top}\beta_j) > \epsilon\} =  \sum_{j=1}^p \pr\{\alpha_j+ x_i^{\top}\beta_j> \Phi^{-1}(\epsilon)\},
\end{equation*}
and thus the goal is computing the following limit
\begin{equation*}
    \lim_{p\to \infty} p \bigg\{1- \Phi \bigg(\frac{\Phi^{-1}(\epsilon) - \mu_p - x_i^{\top}\gamma}{(\tau_p^2 + x_i^{\top}\Psi x_i)^{1/2}}\bigg) \bigg\}.
\end{equation*}
Leveraging the same asymptotic expansion for the
 cumulative distribution function of a normal
distribution used in the proof of Theorem 2,
the above equation is equivalent to 
\begin{equation*}
    \lim_{p \rightarrow \infty} \frac{p (\tau_p^2+ x_i^{\top}\Psi x_i)^{1/2}}{(2\pi)^{1/2}(\epsilon^*-\mu_p-x_i^{\top}\nu)} \exp\bigg\{-\frac{(\epsilon^*-\mu_p-x_i^{\top}\nu)^2}{2(\tau_p^2+ x_i^{\top}\Psi x_i)}\bigg\},
\end{equation*}
with $\epsilon^* = \Phi^{-1}(\epsilon)$. The above equation is the same as equation \eqref{eq:lim_nireg} in the proof of Theorem 2, except for the additive constant $\epsilon^*$. Hence, by simply adapting the same arguments discussed in the aforementioned proof we obtain
\begin{equation*}
    \lim_{p \rightarrow \infty}  E(c_{i}^{\epsilon}) = \gamma  \exp\big\{-\epsilon^* +
        x_i^{\top}\nu +
        (1/2)x_i^{\top}\Psi x_i  -0.5\big\}. 
\end{equation*}

\subsection{Proof of Theorem 3}
Let $w_j=\mathds{1}(n_{+j}>0)$, with $n_{+j} = \sum_{i=1}^n y_{ij}$, be binary random variables that indicate if column $j$ is non empty. Then, the number of non-empty columns is $p^*_n=\sum_{j=1}^p w_j$, where $w_j \sim \mathrm{Bernoulli}[1-\{1-\Phi(\alpha_j)\}^n]$ for the no covariates \texttt{TRACE} model. Therefore, the aim is to compute
\begin{align*}
    \lim_{p \to \infty} E(p^*_n) &= \lim_{p \to \infty} \sum_{j=1}^p E[1-\{1-\Phi(\alpha_j)\}^n]\\
    &= \lim_{p \to \infty} p \int_{-\infty}^{\infty} [1-\{1-\Phi(x)\}^n]\frac{1}{\tau_p} \phi\left(\frac{x-\mu_p}{\tau_p}\right) dx,
\end{align*}
and by  using the substitution $t=\Phi(x)$ the latter equation becomes
\begin{equation}\label{eq:SI_pstarObj}
    \lim_{p \to \infty} p \int_0^1 \{1-(1-t)^n\}\frac{1}{\tau_p} \phi\bigg\{\frac{\Phi^{-1}(t)-\mu_p}{\tau_p}\bigg\} \frac{1}{\phi\{\Phi^{-1}(t)\}}dt.
\end{equation}
By the dominated convergence theorem, we first compute the limit of the integrand in $\eqref{eq:SI_pstarObj}$ and then do the integration. By leveraging the same asymptotic expansion for the inverse of the
cumulative distribution function of a normal distribution used in the proof of Proposition 1 for $\mu_p$, as $p \to \infty$ we obtain 
\begin{equation*}
    \lim_{p \to \infty} p \frac{1}{\tau_p} \phi\bigg\{\frac{\Phi^{-1}(t)-\mu_p}{\tau_p}\bigg\} = \gamma \exp\{-1/2-\Phi^{-1}(t)\},
\end{equation*}
\begin{equation*}
     \lim_{p \to \infty} E(p^*_n) = \gamma \exp(-1/2) \sqrt{2\pi}C_n, \quad C_n = \int_0^1 \{1-(1-t)^n\}\frac{\exp\{-\Phi^{-1}(t)\}}{\exp[-0.5\{\Phi^{-1}(t)\}^2]} dt.
\end{equation*}

If we further assume $\Sigma = I$, then  $p^*_n$ is a sum of independent Bernoulli random variables, and thus $p^*_n \sim \mathrm{Binomial}(p,  \gamma/(\gamma+p) \exp(-1/2) \sqrt{2\pi}C_n)$.
Given that $p\gamma/(\gamma+p) \exp(-1/2) \sqrt{2\pi}C_n$ converges to $\gamma\exp(-1/2) \sqrt{2\pi}C_n$ for $p \to \infty$,  it follows by the law of rare events (see for instance page 113 of \cite{papoulis2002probability}) that $p^*_n$ converges in distribution to a Poisson random variable with parameter $\gamma\exp(-1/2) \sqrt{2\pi}C_n$.

\section{Additional plots for model properties}

\begin{figure}[h]
\begin{center}
    \includegraphics[width=0.67\textwidth]{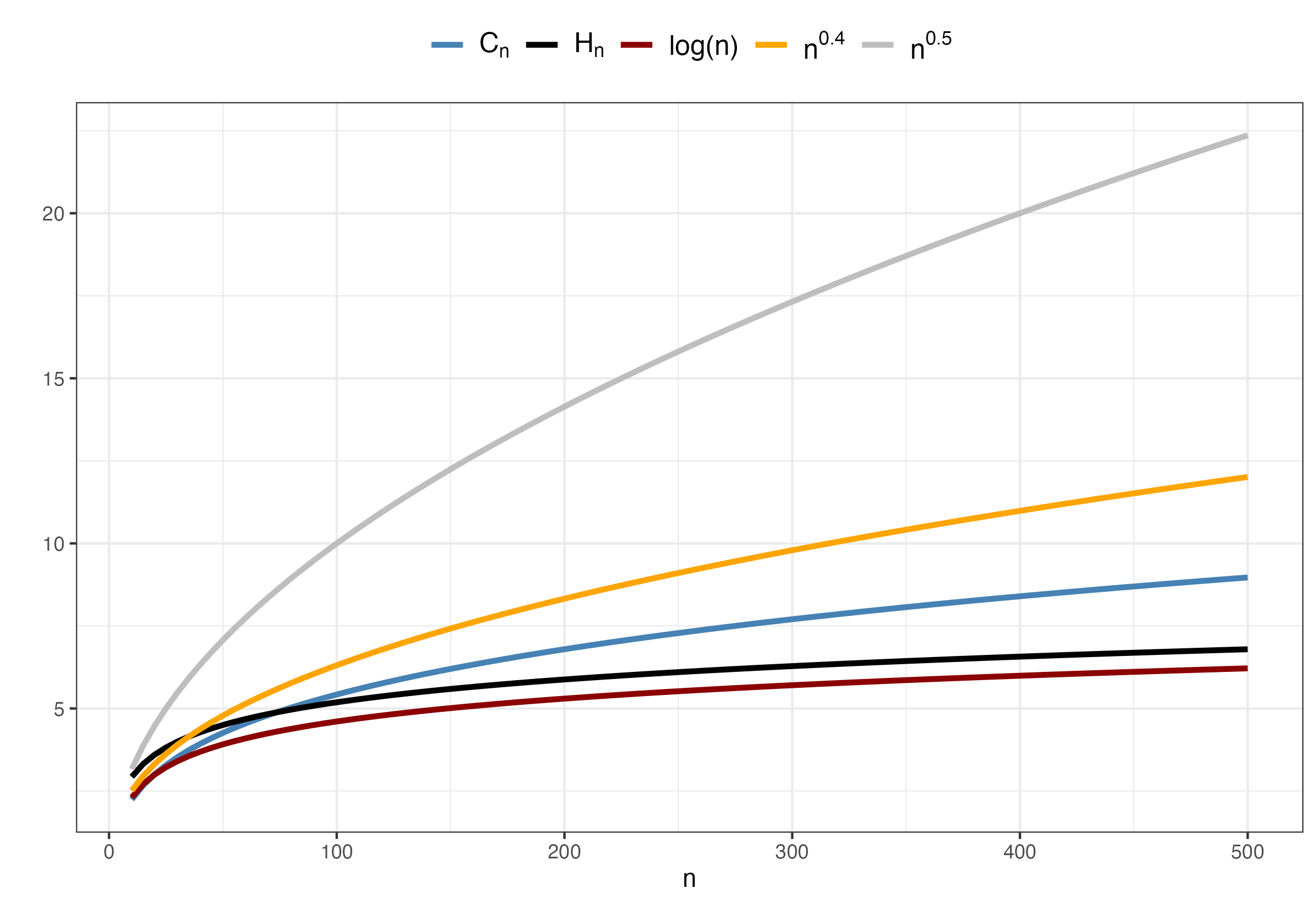}
  \caption{Comparison of $C_n$ defined in Theorem 3 in the main paper, $H_n$ ($n$th harmonic number), $\log(n)$, $n^{0.4}$ and $n^{0.5}$ for different values of $n$.}
    \label{fig:Cn}
    \end{center}
\end{figure}

\section{Posterior computation} \label{sec:sup_computations}

In this section, we describe an algorithm for posterior inference under the \texttt{TRACE} model based on recent computational developments for multivariate probit models \citep{chakraborty2022bayesian}. They developed a two-stage method based on Laplace approximations, which achieves optimal rates for estimating both regression parameters and correlation coefficients. For \texttt{TRACE} we focus on approximating marginal posterior distributions for the parameters, obtaining accurate point estimation, uncertainty quantification, and predictions.

Let $b_j = (\alpha_j, \beta_j)^\top$ denote the random intercept $\alpha_j$ and the regression coefficients $\beta_j$ for species $j$. We first obtain approximations $\Pi_j^*(b_j\mid Y, X)$ for the marginal posteriors of the random intercepts and regression coefficients for $j =1,\dots,p$; we use a simplified likelihood that replaces $\Sigma$ with the identity matrix and apply Laplace approximations. Let $l_j(b_j)$ denote the log-likelihood of the $j$th binary outcome under this simplified model and $\Pi_j(b_j)$ denote the prior distribution for $b_j$, with $\alpha_j \sim N(\mu_p, \tau_p^2)$, $\beta_j \sim N_q(\nu, \Psi)$ and no correlation between the $\alpha_j$s and the $\beta_j$s. 
Then $\Pi_j^*(b_j\mid Y, X) \approx N(\hat{b}_j, Q_j)$, where $\hat{b}_j = \mathrm{argmax}_{b_j}\{l_j(b_j)+\log\Pi_j(b_j)\}$ and $Q_j$ is the corresponding inverse Hessian.
\cite{chakraborty2022bayesian} showed in their Theorem 3.5 that $\Pi_j^*(b_j\mid Y, X)$ is asymptotically normal centered at a maximum composite likelihood estimator with the appropriate variance. 

\begin{figure}[t]
\begin{center}
    \includegraphics[width=0.7\textwidth]{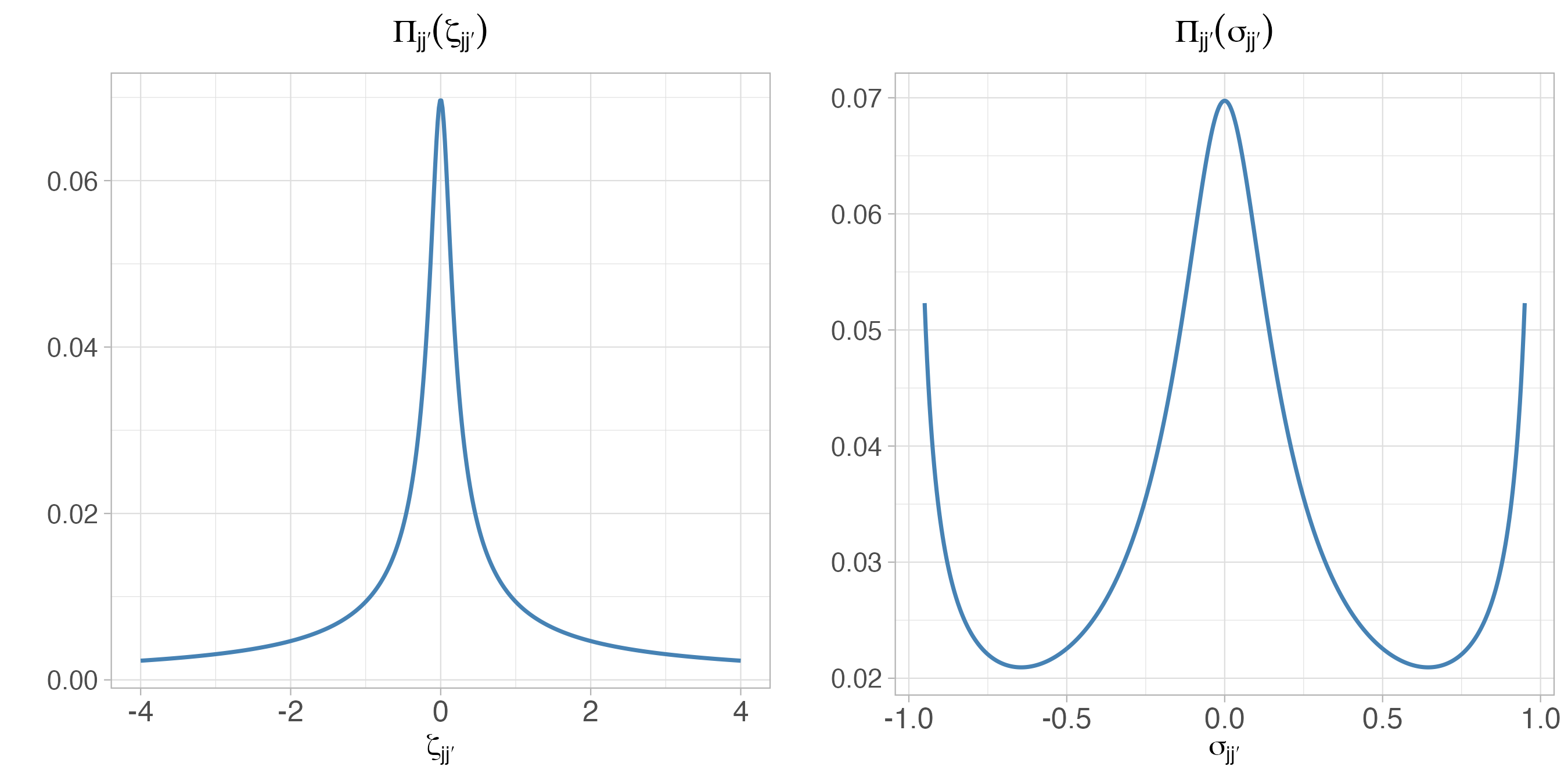}
  \caption{Densities of marginal prior distribution for $\zeta_{jj'}$ in equation (6) in the main paper (left panel) and for $\sigma_{jj'}$ in equation \eqref{eq:SMsigmaPrior} (right panel).}
    \label{fig:sigmaprior}
    \end{center}
\end{figure}

We obtain approximations of the marginal posterior distributions 
$\Pi(\sigma_{jj'} \mid Y,X)$ of the correlations between pairs of outcomes, because the joint
$\Pi(\Sigma \mid Y,X)$ is difficult to interpret and approximate. This is accomplished by considering the likelihood of a bivariate probit model between the pairs $(j,j')$, together with the prior for $\sigma_{jj'}$. We obtain a Gaussian approximation $\Pi^*_{jj'}(\sigma_{jj'} \mid Y,X)$ for the posterior of $\sigma_{jj'}$, as justified in Theorem 3.7 of \cite{chakraborty2022bayesian}. 
The prior $\Pi_{jj'}(\sigma_{jj'})$ on $\sigma_{jj'}$ under $\Pi(\Sigma)$ is available in closed form for several popular classes of priors for correlation matrices. For example, when $\Pi(\Sigma) \propto |\Sigma|^{\nu-1}$ (LKJ), then $\Pi_{jj'}\{(\sigma_{jj'}+1)/2\} \sim \mathrm{Beta}(\nu+p/2-1, \nu+p/2-1)$.
Starting from the hierarchical prior defined in (6) in the main paper, we can derive the marginal prior for $\sigma_{jj'}$.
Integrating out $\omega^2$, the marginal distribution of $\zeta_{jj'}$ is a $t$-distribution with mean 0, scale $b_{\omega}/a_{\omega}$ and degrees of freedom $2a_{\omega}$.
By applying a change of variable, we can obtain the marginal prior for $\sigma_{jj'}$
\begin{equation} \label{eq:SMsigmaPrior}
     \Pi_{jj'}(\sigma_{jj'}) = \frac{\Gamma(a_\omega+1/2)}{(2\pi b_{\omega})^{1/2}\Gamma(a_\omega)}  \bigg[ 1+ \frac{1}{2b_{\omega}} \bigg\{ \frac{1}{2}\log\bigg(\frac{1+\sigma_{jj'}}{1-\sigma_{jj'}}\bigg)\bigg\}^2\bigg]^{-(a_{\omega}+1/2)} \frac{1}{1-\sigma_{jj'}^2},
\end{equation}
where $\Gamma(\cdot)$ is the gamma function. Figure \ref{fig:sigmaprior} shows the marginal prior distributions $\Pi_{jj'}(\zeta_{jj'})$  and $\Pi_{jj'}(\sigma_{jj'})$ with $a_{\omega} = b_{\omega} = 0.01$, the hyperparameter values we used in the simulations and fungi application.

The mean $\hat{\sigma}_{jj'}$ and the variance $s_{jj'}^2$ of the approximate posteriors for $\sigma_{jj'}$ are obtained using Gauss-Legendre quadrature.
\cite{chakraborty2022bayesian} show that 
asymptotically the distribution is centered at the maximum composite likelihood estimator from the bivariate margins. 
In the common correlation matrix setting, assuming  $\rho_{jj'} = \rho$, we use Metropolis-Hastings to sample from the conditional posterior of $\rho$ given the $b_j$s; for this simple choice of correlation structure, we do not use the above asymptotic Gaussian approximation for the posterior of $\sigma_{jj'}$. Predictions can be obtained using pairwise approximations to the posterior predictive.

The above approximations are defined conditionally on the hyperparameters $\gamma$ and $\omega^2$. As estimates for $\gamma$ and $\omega^2$, we use the means of samples from the approximate Gibbs sampler described in Algorithm \ref{algo:composite}. In the case of a common correlation coefficient, the algorithm can be simplified as we do not need to estimate $\omega$; therefore, we only repeat steps 1 to 3.

\begin{algorithm}[h] \label{algo:composite}

   \caption{Approximate Gibbs sampler to estimate hyperparameters for the \texttt{TRACE} model with a hierarchical shrinkage prior on the correlation coefficients.}
   Initialize $\omega$ and $\gamma$.\\
   \vspace{-0.1cm}
   \begin{enumerate}
    \item  Given $\gamma$ obtain approximation to $\Pi(\alpha_j\mid Y, X, \gamma)$ as $N(\hat{\alpha}_j, Q_j)$ for $j=1, \dots,p$ with the prior $\alpha_j \sim N(\mu_p, \tau_p^2)$. 
    \item  Draw $\alpha_j \sim N(\hat{\alpha}_j, Q_j)$ independently for $j=1,\dots,p$.\\
    \item Update $\gamma$ through a Metropolis–Hastings step.
    \item Given $\omega$, obtain approximations to $\Pi(\sigma_{jj'} \mid Y, X, \omega)$ as $N(\hat{\sigma}_{jj'}, s_{jk}^2)$ for $j<j'=1, \dots,p$.
     \item Draw $\sigma_{jj'}\sim N(\hat{\sigma}_{jj'}, s_{jk}^2)$ independently and set $\zeta_{jj'}=0.5 \log \{(1+\sigma_{jj'})/(1-\sigma_{jj'})\}$\\
     for $j<j'=1, \dots,p$.
     \item Update $\omega^2 \sim IG\left\{\frac{p(p-1)}{2}+a_{\omega}, \frac{1}{2}\sum_{j<j'} \zeta_{jj'}^2 + b_{\omega} \right\}$.
   \end{enumerate}
 Repeat steps 1-6 T times to obtain T samples of $\omega$ and $\gamma$.\\
\end{algorithm}

The same approach described in Algorithm \ref{algo:composite} can  be applied to incorporate the hierarchical prior on the regression coefficients, defined in equation \eqref{eq:b_hier} in the main paper. In this case, after have obtained an approximation of $\Pi(\beta_j\mid y, X, \nu, \Psi)$ with prior $\beta_j \sim N_q(\nu, \Psi)$ and having drawn a sample from the corresponding posterior of $\beta_j$,  one can update $(\nu, \Psi)$ as $(\nu, \Psi) \sim \mathrm{NIW}(\nu_p, \iota_p, d_p, \Xi_p)$, with $\iota_p = \iota + p$, $d_p = d +  p$, $\nu_p = (\iota\nu_0 + p\hat{\beta})/\iota_p$ with $\hat{\beta} = (\sum_{j=1}^p \beta_j)/p$ and $\Xi_p = \Xi + S + (p\iota/\iota_p)\sum_{j=1}^p (\beta_j - \nu_0)(\beta_j - \nu_0)^{\top}$ with $S = \sum_{j=1}^p (\beta_j - \hat{\beta})(\beta_j - \hat{\beta})^{\top}$.

As an alternative to the above composite-likelihood and Laplace-based posterior approximations, we additionally ran a wide variety of experiments using data augmentation Markov chain Monte Carlo algorithms targeting the exact posterior distribution. Such algorithms were not competitive with the above approach in terms of computational time or estimation accuracy.

\section{Additional simulations} \label{sec:suppSim}

\begin{figure}[t]
\begin{center}
    \includegraphics[width=0.7\textwidth]{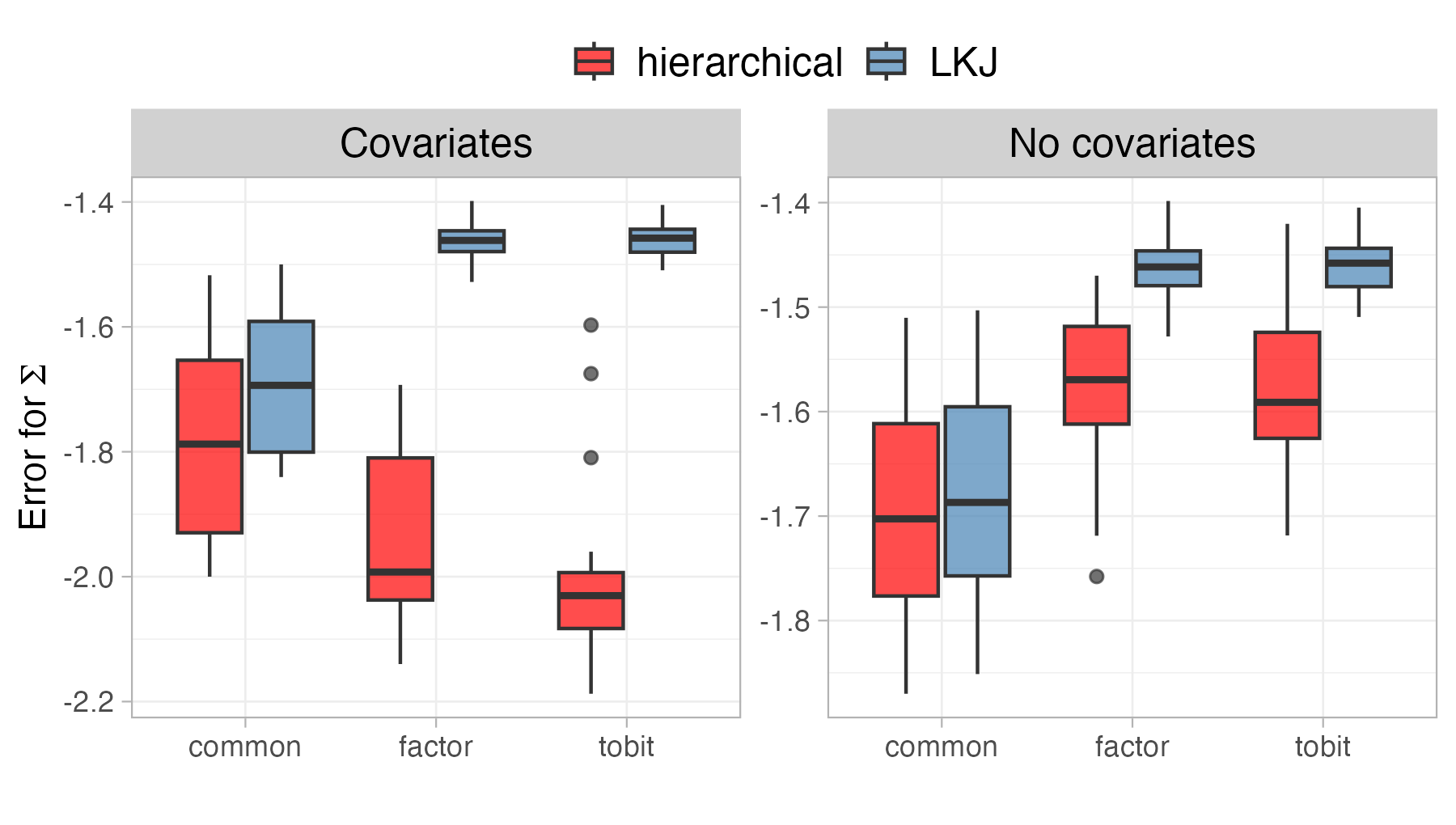}
    \caption{Logarithm of Frobenius errors for $\Sigma^*$ of the \texttt{TRACE} model with \texttt{LKJ} and hierarchical priors for $\sigma_{jj'}$ in 20 simulations for each scenario without (left) and with covariates (right).}
    \label{fig:lkj}
    \end{center}
\end{figure}

In this section, we report results from additional simulation experiments.
We start by showing the improved performance for the \texttt{TRACE} model with the hierarchical prior for $\sigma_{jj'}$s defined in \eqref{eq:corrH} of the main article relative to the \texttt{LKJ} prior \citep{lewandowski2009generating}. We considered the same simulation settings described in Section \ref{sec:simulation} of the main paper and we set the \texttt{LKJ} parameter to one. Figure \ref{fig:lkj} shows the performance in estimating the elements of $\Sigma^*$. Although the accuracy in estimating  $\pi^*$ and  $\Delta^*_{100|80}$ was comparable between the two priors, the hierarchical correlation prior had consistently better performance in estimating the correlations in all scenarios, including settings with and without covariates. 

\begin{figure}[t]
\begin{center}
\includegraphics[width=0.75\textwidth]{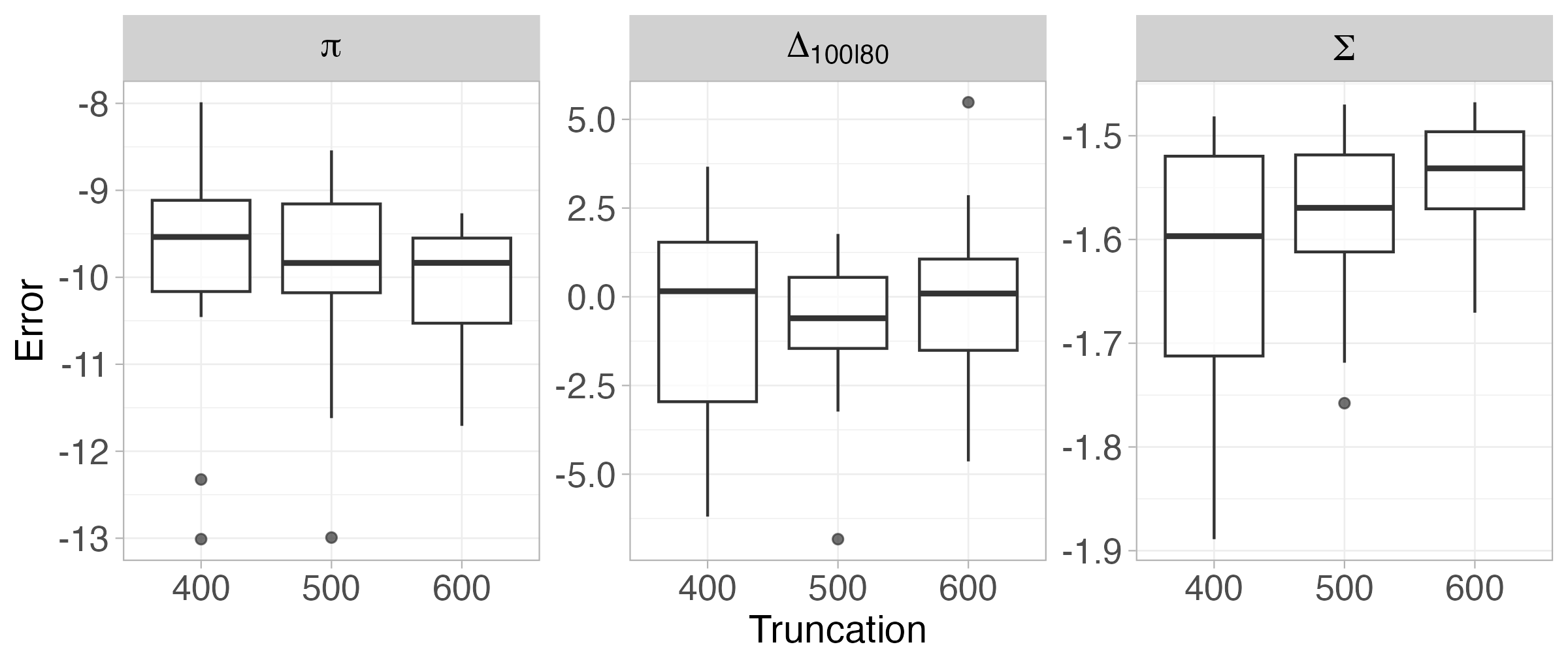}
  \caption{Logarithm of Frobenius errors for $\pi^*$, $\Delta_{100|80}^*$, and $\Sigma^*$ of the \texttt{TRACE} model in 20 simulations for different truncation levels.}
    \label{fig:truncation}
    \end{center}
\end{figure}

We provide simulations demonstrating that posterior inferences for the \texttt{TRACE} model are robust to the choice of truncation level for the number of species. We simulate 20 datasets with the same values for $\gamma$ and $n$ used in Section \ref{sec:simulation} of the main paper, and consider three different truncation levels for the number of species $p \in \{400,500,600\}$ for each of the 20 simulations. We evaluated the performance of \texttt{TRACE} in estimating $\Sigma^*$, $\pi^*$ and $\Delta_{100|80}^*$ using the same metrics described in Section \ref{sec:simulation} of the main article. For illustration purposes, we show the results for the `factor' scenario without covariates, but the conclusions were comparable across all scenarios and for settings with and without covariates. The results, displayed in Figure \ref{fig:truncation}, show that the error metrics for all quantities of interest are of the same order at the different truncation levels.

\section{Additional plots for fungal biodiversity application}

\begin{figure}[h]
\begin{center}
\includegraphics[width=0.97\textwidth]{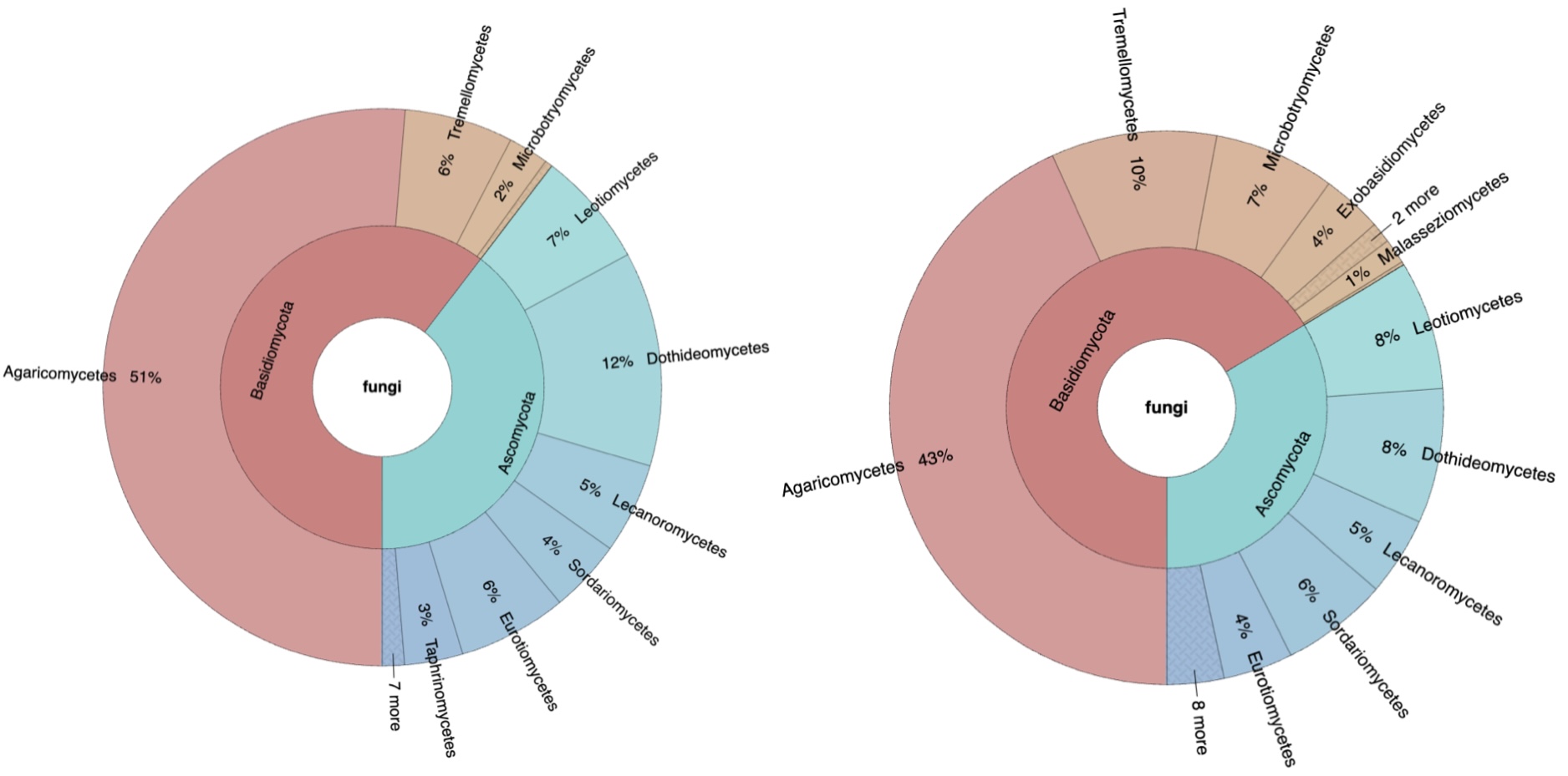}
  \caption{Fungal taxonomic composition and marginal species occurrence probabilities for Site 3 (left) and Site 4 (right).}
\label{fig:Krona_supp}
\end{center}
\end{figure}

\begin{figure}[h]
\begin{center}
    \includegraphics[width=0.7\textwidth]{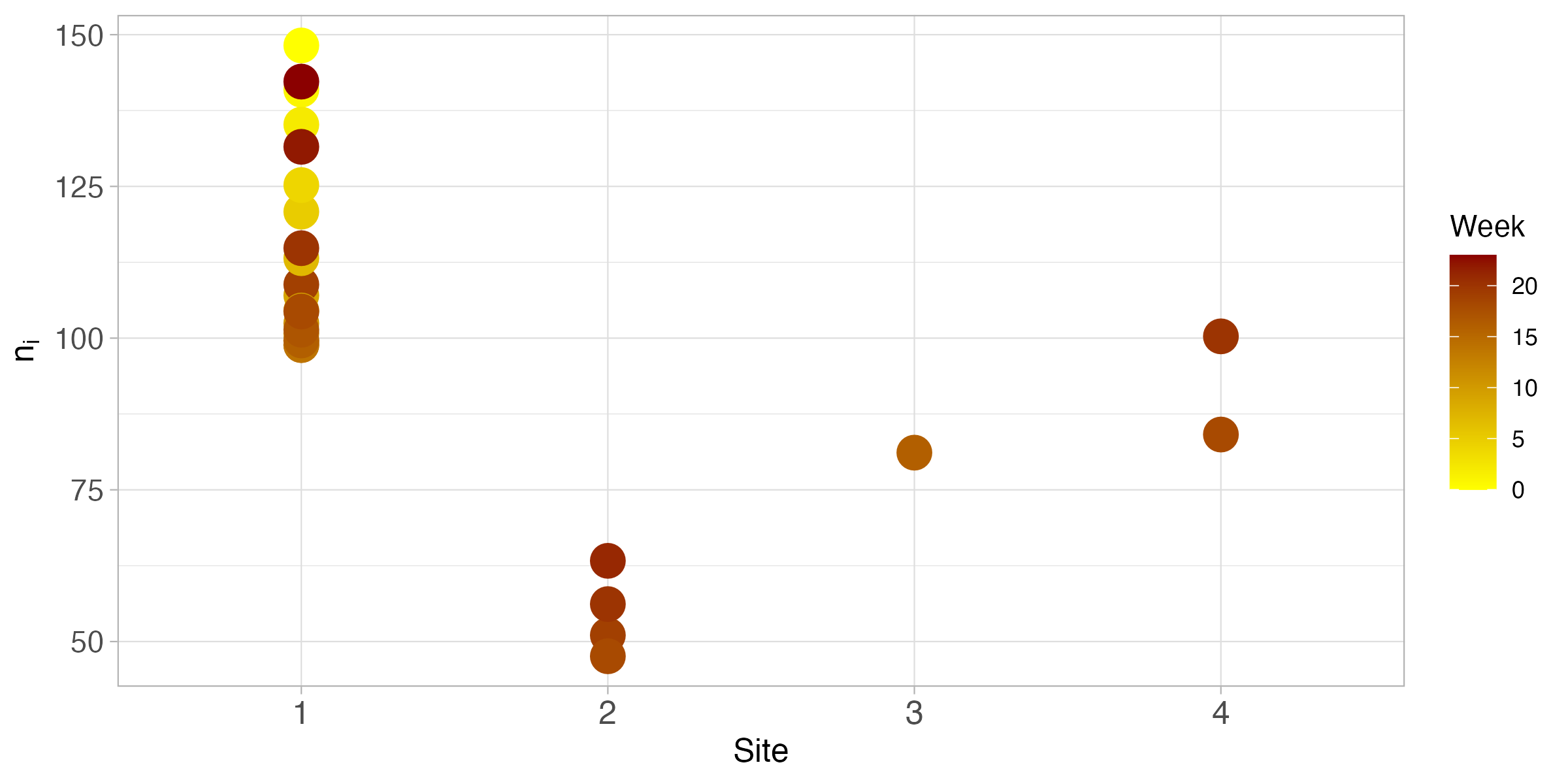} 
\caption{Posterior mean of sample-specific species richness for the four sites in different weeks.}
    \label{fig:spSite}
\end{center}
\end{figure}

Figure \ref{fig:Krona_supp} shows a snapshot of the Krona wheel plot specific to Site 3 and Site 4, highlighting the differences in the taxonomic composition of the fungi at the different sites. Figure \ref{fig:spSite} shows the posterior means of sample-specific species richness in the four different locations in Finland in different weeks. The week values cover the entire fruiting period, from early spring (week 0) to late autumn (week 22). In general, site 1, located in central Finland, exhibits a greater species richness compared to the other ones. This site shows  seasonal variation in community composition, with a decreasing trend in species richness in late spring and summer, as observed in  \cite{AbregoFungi}.

\end{document}